\documentclass{article}
\usepackage[table, svgnames, dvipsnames]{xcolor}
\usepackage[margin=1in]{geometry}
\usepackage{graphicx}
\usepackage{float}
\usepackage{wrapfig}
\usepackage{soul}
\usepackage{url}

\usepackage{booktabs}
\usepackage{multirow}
\usepackage{amsmath}

\usepackage{amsmath,amssymb}

\usepackage{tikz}
\usetikzlibrary{arrows.meta,decorations.pathreplacing,positioning}

\usepackage{longtable}
\usepackage{makecell, cellspace, caption}
\usepackage{array}
\newcolumntype{L}[1]{>{\raggedright\let\newline\\\arraybackslash\hspace{0pt}}m{#1}}
\newcolumntype{C}[1]{>{\centering\let\newline\\\arraybackslash\hspace{0pt}}m{#1}}
\newcolumntype{R}[1]{>{\raggedleft\let\newline\\\arraybackslash\hspace{0pt}}m{#1}}

\title{Building Envelope Inversion by Data-driven Interpretation of Ground Penetrating Radar}
 
\author{
Ahmed Nirjhar Alam$^1$ \and
Wesley F. Reinhart$^{1,2}$\thanks{Address correspondence to: reinhart@psu.edu} \and
Rebecca K. Napolitano$^3$
}

\date{
	$^1$Department of Materials Science and Engineering, The Pennsylvania State University \\ %
    $^2$Institute for Computational and Data Sciences, The Pennsylvania State University \\
	$^3$Department of Architectural Engineering, The Pennsylvania State University\\[2ex]%
}

\begin{document}

\maketitle

\section{Abstract}

Ground-penetrating radar (GPR) combines depth resolution, non-destructive operation, and broad material sensitivity, yet it has seen limited use for diagnosing building envelopes. The compact geometry of wall assemblies, where reflections from closely spaced studs, sheathing, and cladding strongly overlap, has made systematic inversion of GPR data difficult. Recent progress in data-driven interpretation provides new opportunities to revisit this challenge and assess whether machine learning can reliably extract structural information from such complex signals.

Here, we develop a data-driven GPR-based inversion framework that decomposes wall diagnostics into classification tasks addressing vertical (stud presence) and lateral (wall-type) variations. Alongside model development, we implement multiple feature minimization strategies -including recursive elimination, agglomerative clustering, and L$_0$-based sparsity -to promote fidelity and interpretability. Among these approaches, the L$_0$ based sparse neural network (SparseNN) emerges as particularly effective: it exceeds Random Forest accuracy while relying on a fraction of the input features, each linked to identifiable dielectric interfaces. SHAP analysis further confirms that the SparseNN learns reflection patterns consistent with physical layer boundaries.

In summary, this framework establishes a foundation for physically interpretable and data-efficient inversion of wall assemblies using GPR radargrams. Although defect detection is not directly addressed, the ability to reconstruct the intact envelope and isolate features tied to structural elements provides a necessary baseline for future defect localization and condition assessment.

\section{Introduction}
Building envelope, the interface between indoor and outdoor environments, plays a central role in regulating heat, air, and moisture flows. Its performance directly affects energy efficiency, occupant comfort, and durability. Defects such as moisture intrusion, air leakage, and insulation failures can accelerate material deterioration, lead to microbial growth, and, in turn, result in substantial maintenance costs, reduced asset value, and adverse respiratory health impacts for occupants~\cite{who_dampness_2009, mendell_2011, Peuhkuri2003}.

Accurate diagnostics are essential to identify and address envelope deficiencies before they cause irreversible damage. In many diagnostic workflows, characterizing the construction of wall assemblies is a vital precursor: interpretation of defect indicators depends heavily on prior knowledge about underlying material composition and configuration~\cite{astm_c1060, astm_d6432}. Wall characterization is also an important objective in its own right, directly supporting renovation planning, verification of code compliance, conservation of historical structures, preparation of material inventories for retrofits, and identification of hazardous materials before demolition or renovation.~\cite{ufgs_becx, ashrae_211, historic_england, wong_heritage_ndt, epa_neshap}. Practically, the problem of characterizing envelope components can be viewed as an as-built assembly inversion problem: recovering the wall’s internal structure from observational data, even when no visible defects are present.

Non-destructive evaluation (NDE) methods offer a more attractive paradigm than destructive inspections by enabling in-situ, repeatable, and large-area characterization without compromising structural integrity~\cite{malhotra2004}. 
Compared to intrusive sampling, which is costly and localized, NDE approaches can rapidly screen entire wall sections to guide targeted interventions.
NDE methods applied to building envelope assessment include infrared thermography (IRT) and ultrasonic testing (UT), which are widely used, as well as ground-penetrating radar (GPR), which is increasingly employed where subsurface or multi-layer characterization is required \cite{ISO6781, ASTM_C597, Gilmutdinov2022}.
IRT allows rapid surface temperature mapping but is highly sensitive to environmental conditions such as ambient temperature, wind, and solar exposure~\cite{kylili2014_irt}.
UT offers high resolution in defect detection and thickness measurement, but typically requires direct contact and couplant, limiting its scalability for large wall areas~\cite{aci_2282r, nist_ir_7974}.

GPR presents a compelling alternative: it is non-contact, portable, capable of covering large areas quickly, and offers meaningful penetration depth in common building materials.
Its effectiveness does not depend on transient environmental differences, unlike IRT, and it avoids surface prep or coupling requirements that limit UT~\cite{solla_transport_review, Benedetto2017_SignalProcessing}. 
These characteristics position GPR as particularly suitable for building envelope assessment.

The interpretation of GPR data, however, is inherently complex \cite{Daniels2004,Lombardi2022}. Moreover, the interpretation issues are particularly severe for building envelopes. Such assemblies often contain thin layers whose thickness is near or below the emitted wavelength, leading to weak or ambiguous reflections \cite{Hosseini2014, zhao2015_thinlayers}.
The low dielectric contrast between adjacent layers further diminishes signal amplitude, while structural heterogeneity introduces multiple scattering and higher-order echoes that can be confused with genuine material boundaries~\cite{hugenschmidt2010_concrete, lahouar2008_layers, oliviera2021_svdclutter}.
Due to such interpretability difficulties, the GPR has not seen widespread use in building envelope diagnostics compared to other NDE modalities.

Over the past decade, data-driven approaches, especially machine learning, have greatly advanced GPR interpretation in areas such as subsurface utility mapping, pavement structure analysis, and archaeological investigations~\cite{bai2023_gpr_ml_review, gprinvnet2021, dmrf_unet2022, Aziz2024_GPR_BayesianInversion,Aziz2025_GPR_FDTD}.
Such methods automate feature extraction, improve classification performance, and reduce subjectivity, making them a logical choice for building envelope characterization.
Recent work shows that learning-based models can achieve robust classification in complex environments, minimizing dependence on expert-crafted rules and improving reproducibility across sites~\cite{bai2023_gpr_ml_review, hou2022_review, gprinvnet2021, dmrf_unet2022}. 

In the context of these advancements, the application of GPR to building envelopes is worth revisiting via the use of data-driven signal interpretation. 
However, despite this promise, data-driven methods present notable challenges that must be addressed. 
For instance, a common drawback of data-driven methods is the lack of interpretability (the so-called ``black box'' problem). 
In engineering diagnostics, this limitation is particularly critical: high-stakes decisions - such as intrusive investigation, abatement, or retrofit, require models whose predictions can be explained and audited, rather than accepted as opaque black boxes~\cite{rudin2019}. 
In the context of GPR-based diagnosis of building envelopes, interpretability would entail relating predictive cues in the time-domain signal to physically meaningful wall features, such as material interfaces and structural elements, and ensuring that these relationships are consistent and physically plausible.

Another closely related consideration is feature efficiency, which is particularly relevant for GPR. 
Individual GPR traces are high-dimensional and often contain redundant information; identifying and retaining only the most informative time samples can both reduce computational demands and improve model generalization by discouraging reliance on spurious signal artifacts that fail to transfer across sites or wall configurations.

In this work, we investigate data-driven approaches for the interpretation of GPR data in building envelope diagnostics, with a particular emphasis on model interpretability. 
Rather than attempting direct inversion of GPR scans to wall configurations, we address the problem indirectly through two linked classification tasks: stud detection and wall type identification. 
We begin by establishing performance baselines using widely adopted machine learning models, then analyze the gains in both accuracy and feature efficiency achievable with the strongest of these baselines. 
Building on these insights, we unify the goals of high predictive performance and compact, physically meaningful feature sets within a single framework using sparse neural networks. 
This framework not only achieves superior results but also yields features whose significance and interpretability align closely with underlying wall structures, offering strong support for the approach in diagnostic applications.

\section{Data and Methods}
\subsection{Data}
\begin{figure}[!h]
    \centering
    \includegraphics[width = 1\linewidth]{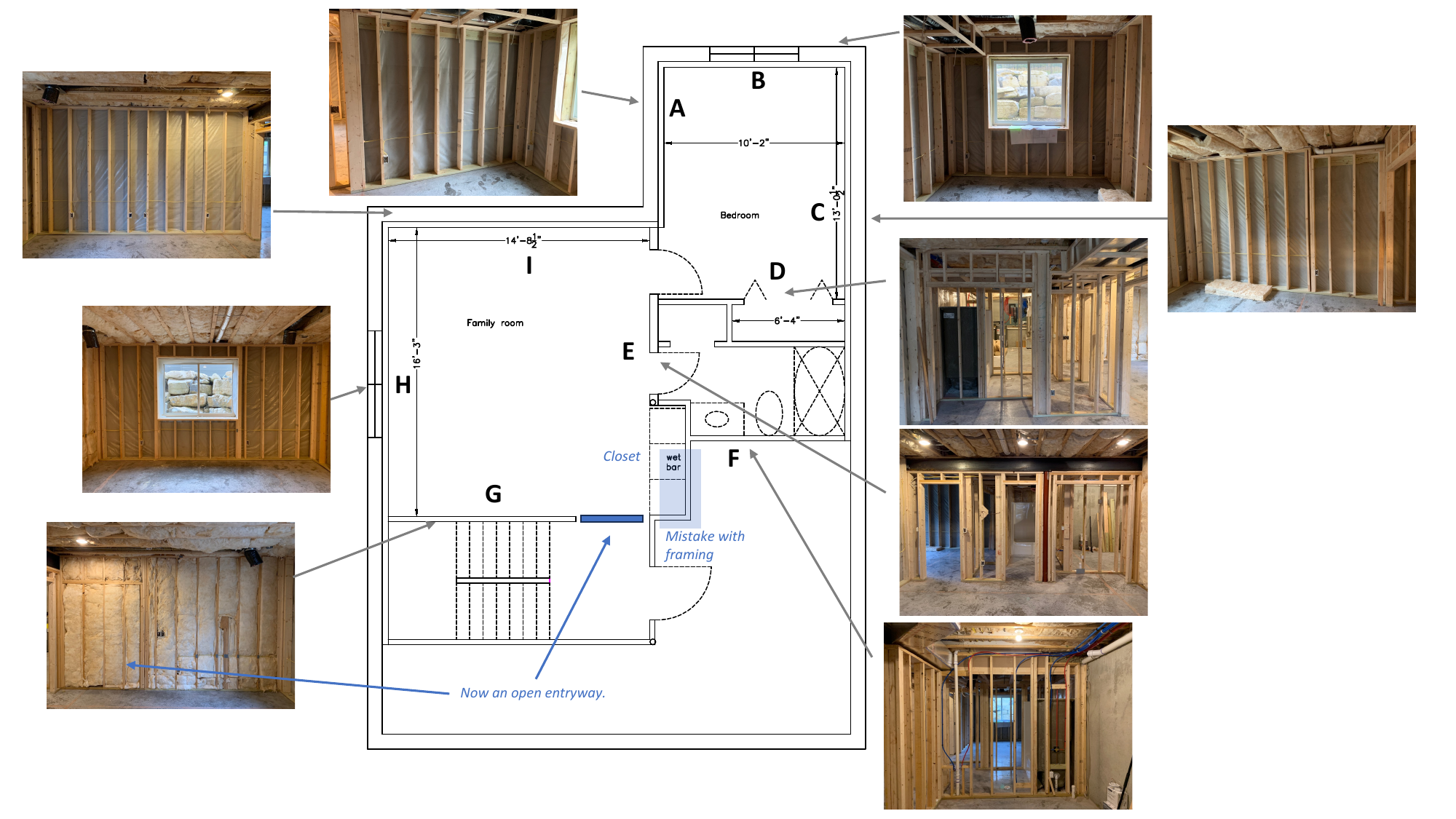}
    \caption{House floor plan with inset images of building envelope during construction.}
    \label{fig:floorplan}
\end{figure}

\begin{figure}[H]
    \centering
    \includegraphics[width = 1\linewidth]{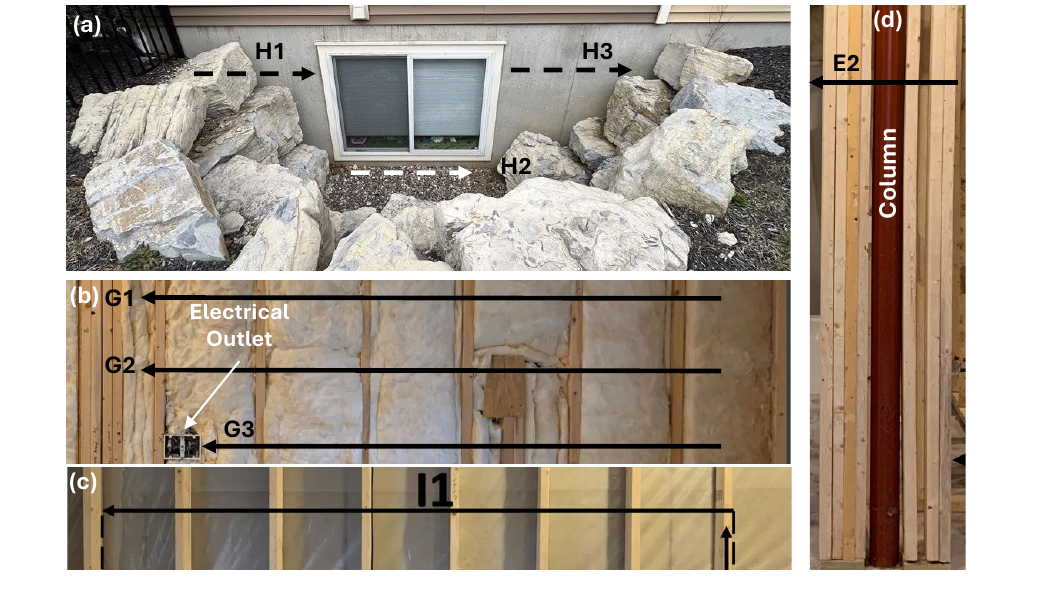}
    \caption{Each wall in the dataset is assigned an alphabetic label, and individual wall segments are indexed numerically; the resulting alphanumeric identifier uniquely specifies each GPR scan. Panel (a) shows exterior wall segments with partial rock background (\texttt{H1} and \texttt{H3}) and soil background (\texttt{H2}). Panel (b) shows interior wall segments, including \texttt{G1}, as well as \texttt{G2} and \texttt{G3}, which are located near electrical outlets. Panel (c) shows an exterior wall segment (\texttt{I1}). Panel (d) shows an interior wall segment (\texttt{E2}) containing an embedded column. Unless otherwise noted, wall segments \texttt{I1} (exterior) and \texttt{G3} (interior) are used in the training set for all classification and interpretability tasks considered in this study, with the exception of the wall-background classification task.}
    \label{fig:wall_variations}
\end{figure}

Field B-scans collected from a two-storied residential building were used as data for the study (Figure~\ref{fig:floorplan}).
All building walls scanned were at the below-grade level of the basement, which is partially buried in the soil (Figure~\ref{fig:wall_variations} 
and Appendix: Figure~\ref{fig:window_well_schematic}). The exterior walls of the building are also surrounded by a semi-circular arrangement of rocks embedded in soil (window wells), as seen in Figure \ref{fig:wall_variations} and Appendix: Figure~\ref{fig:window_well_schematic}. 
This leads to gradually varying backgrounds for such walls. 
Additionally, the proximity to different wall utilities was a variation in interior wall scans (Figure~\ref{fig:wall_variations} - Column and electrical outlet). 
The floor plan of the building, with images of the interior of the relevant building walls, is illustrated in Figure~\ref{fig:floorplan}.

The different B-scans were labeled using an alphanumeric convention where the alphabet corresponded to a particular wall and the numeric value corresponded to a specific segment in the wall. Thus, \texttt{G3} corresponds to wall \texttt{G}'s segment \texttt{3}. (Figure~\ref{fig:wall_variations})

\begin{figure}[H]
    \centering
    \includegraphics[width = 1\linewidth]{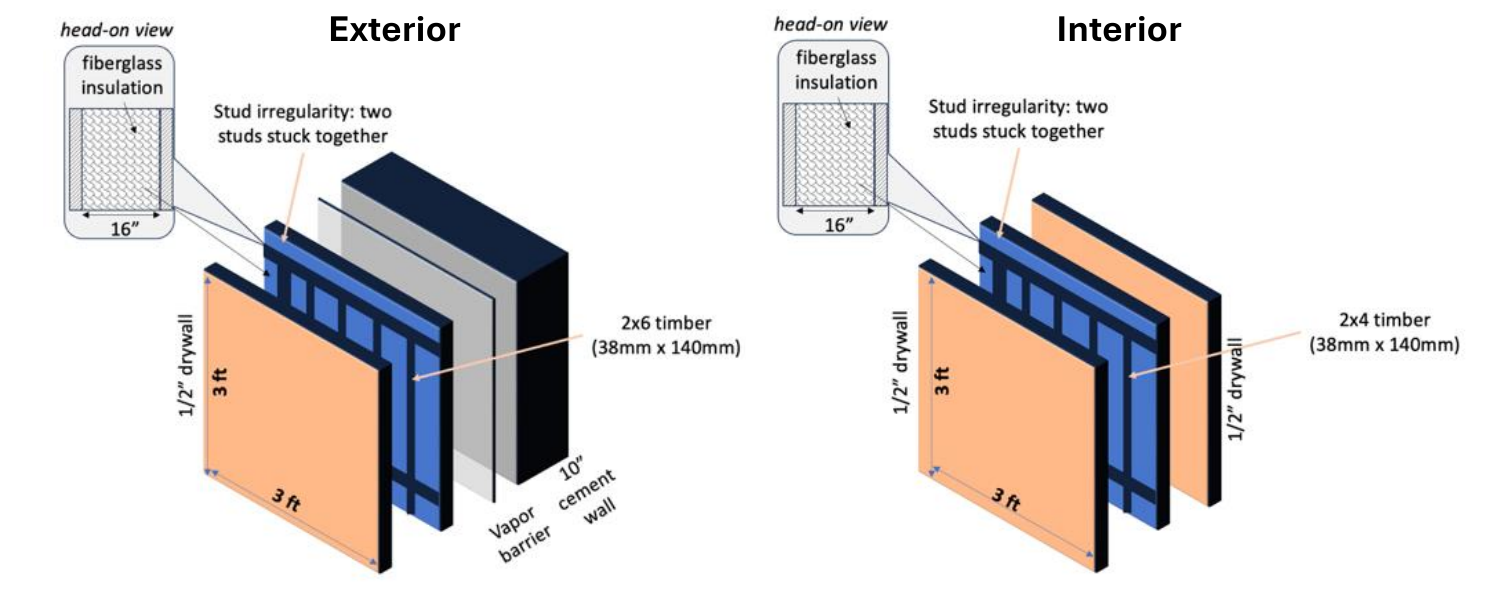}
    \caption{Insulation layers (and stud depth) are four inches for interior walls and six inches for exterior walls. Interior walls are drywalled on both sides. Exterior walls consist of drywall on the interior side (where the scans occur) and cement walls on the exterior. A vapor barrier separates the insulation and cement layers.}
    \label{fig:wall_configurations}
\end{figure}

The dataset comprised building envelopes in two categories: interior and exterior. Their configurations are shown in Figure~\ref{fig:wall_configurations}.

All thickness and spacing values are reported in the 'x' inches ('x' cm) format in the paper. The actual sizes are indicated in inches in brackets for components whose nominal sizes differ from their actual values. 

The interior partition walls are non-load-bearing assemblies framed with \#2 grade Spruce–Pine–Fir (SPF) dimensional lumber. Each wall consists of 1/2'' (1.27 cm) gypsum wallboard (GWB) on both sides, with a 4'' (10.16 cm) cavity filled with R-19 Kraft-faced fiberglass batt insulation for sound attenuation. The framing members are nominal 2''×4'' (5.08 cm × 10.16 cm) studs (actual size 1.5''×3.5'' or 3.81 cm × 8.89 cm) spaced at 16'' (40.64 cm) on-center. Around door and window openings, the framing incorporates built-up stud assemblies comprising king studs, jack studs, and headers to support the load transfer around the openings. 

The exterior foundation walls are composed of several layers, arranged from outside to inside as follows: a 10'' (25.4 cm) poured concrete foundation wall, a 4-mil clear polyethylene vapor barrier applied directly against the concrete, a framed wall using 2''×6'' (5.08 cm × 15.24 cm) Spruce–Pine–Fir lumber (specified as \#2 \& BTR grade, with \#1 grade pressure-treated members at moisture-prone locations), and R-21 Kraft-faced fiberglass batt insulation filling the stud cavities. The interior surface is finished with 1/2'' (1.27 cm) gypsum wallboard. The Kraft facing serves as a secondary, interior vapor retarder. 

Ground-penetrating radar (GPR) scans were collected using the Proceq GP8800 system, a compact, handheld stepped-frequency continuous-wave (SFCW) GPR unit with a modulated frequency range of 400–6000~MHz and a maximum penetration depth of 65~cm \cite{proceqGP8800}. In SFCW operation, a sequence of discrete-frequency sine waves is transmitted, and the received responses are combined via inverse Fourier transform to reconstruct the time-domain signal \cite{daniels}. SFCW offers a high signal-to-noise ratio, broad dynamic range, and strong immunity to radio-frequency interference \cite{NICOLAESCU201259}. Its depth resolution is directly determined by the system bandwidth, with larger frequency sweeps yielding finer vertical separation between resolvable targets \cite{fhwa}—a property advantageous for building envelope diagnostics, where closely spaced layers must be distinguished. A known drawback is its sensitivity to motion during the frequency sweep, which can introduce phase errors and range shifts if not properly controlled \cite{pieraccini}.

Each wall segment was scanned three times in the direction indicated by arrows in the corresponding figures, yielding three samples per segment. Each B-scan trace was 12~ns in duration and comprised 655 uniformly spaced time samples. For modeling and analysis, B-scans were treated as two-dimensional arrays of variable width, with each column corresponding to an individual A-scan (trace). Each trace was considered a single data point in all subsequent analyses.

\subsubsection{Preprocessing}
Preprocessing steps were intentionally kept minimal to evaluate the performance of the proposed data-driven methods under low manual intervention.

A fixed exponential gain with an exponent of 0.8 was applied to the A-scans to enhance signal quality. This transformation amplified lower amplitude values and compressed higher amplitude peaks according to:

\begin{equation} \label{eq:gain_equation}
    A_g = A_0^\gamma
\end{equation}

Here, $A_0$ denotes the normalized A-scan, and the gain factor satisfies $0 < \gamma < 1$. The transformed signal $A_g$ represents the gain-adjusted A-scan.
The value $\gamma = 0.8$ was selected based on empirical evaluations of visual clarity and downstream model performance.

\subsubsection{Dataset Challenges}

Preliminary data analysis revealed several challenges in utilizing the field scans from a data science perspective.

\paragraph{Unpredictable Far-field Variations}\leavevmode\newline

Later signals exhibit systematic variation within the same walls. Raw B-scans primarily show perturbations within the first 2 ns, with very low visibility. Applying linear time-variant gain enhances later signals but does not fully distinguish patterns. Finally, time-variant exponential gain reveals inconsistent patterns appearing later in the signal. 

\begin{figure}[H]
    \centering
    \includegraphics{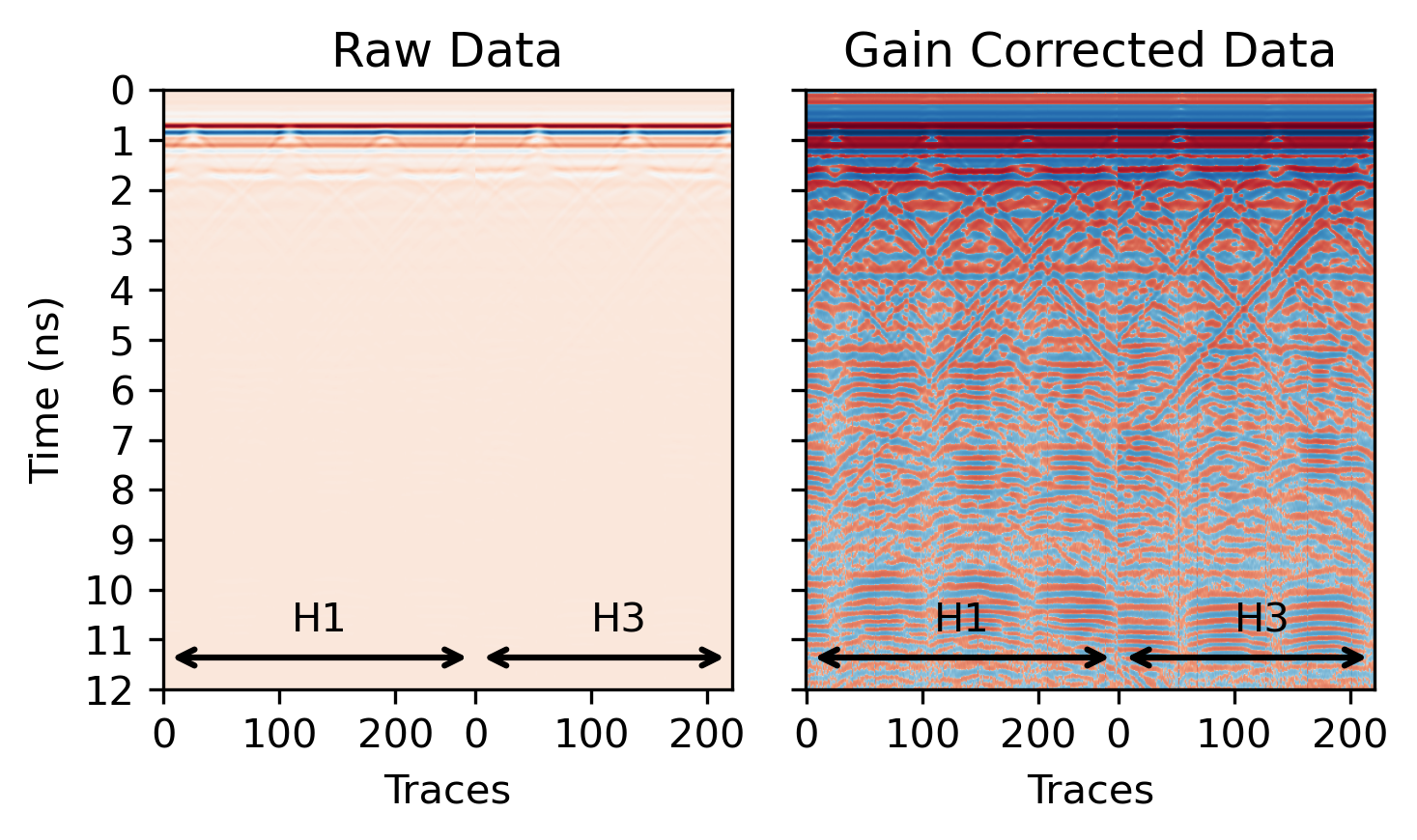}
    \caption{Two scans from the same wall (exponential time invariant gain)}
    \label{fig:wall_comparison}
\end{figure}

As per specification, scans \texttt{H1} and \texttt{H3}, being part of the same exterior wall, have nearly identical cross-sections. They also have the same background at different portions of the wall. However, as seen in Figure~\ref{fig:wall_comparison}, the signals between 10-11 ns show distinct and consistent differences. This may be due to minute differences in the dimensions of envelope components (within the manufacturer's error tolerance) and/or differences in lateral backgrounds, leading to shifts in the signal that increase with time.     


\paragraph{Complexity}\leavevmode\newline

\begin{figure}[H]
    \centering
    \includegraphics[width = 1\linewidth]{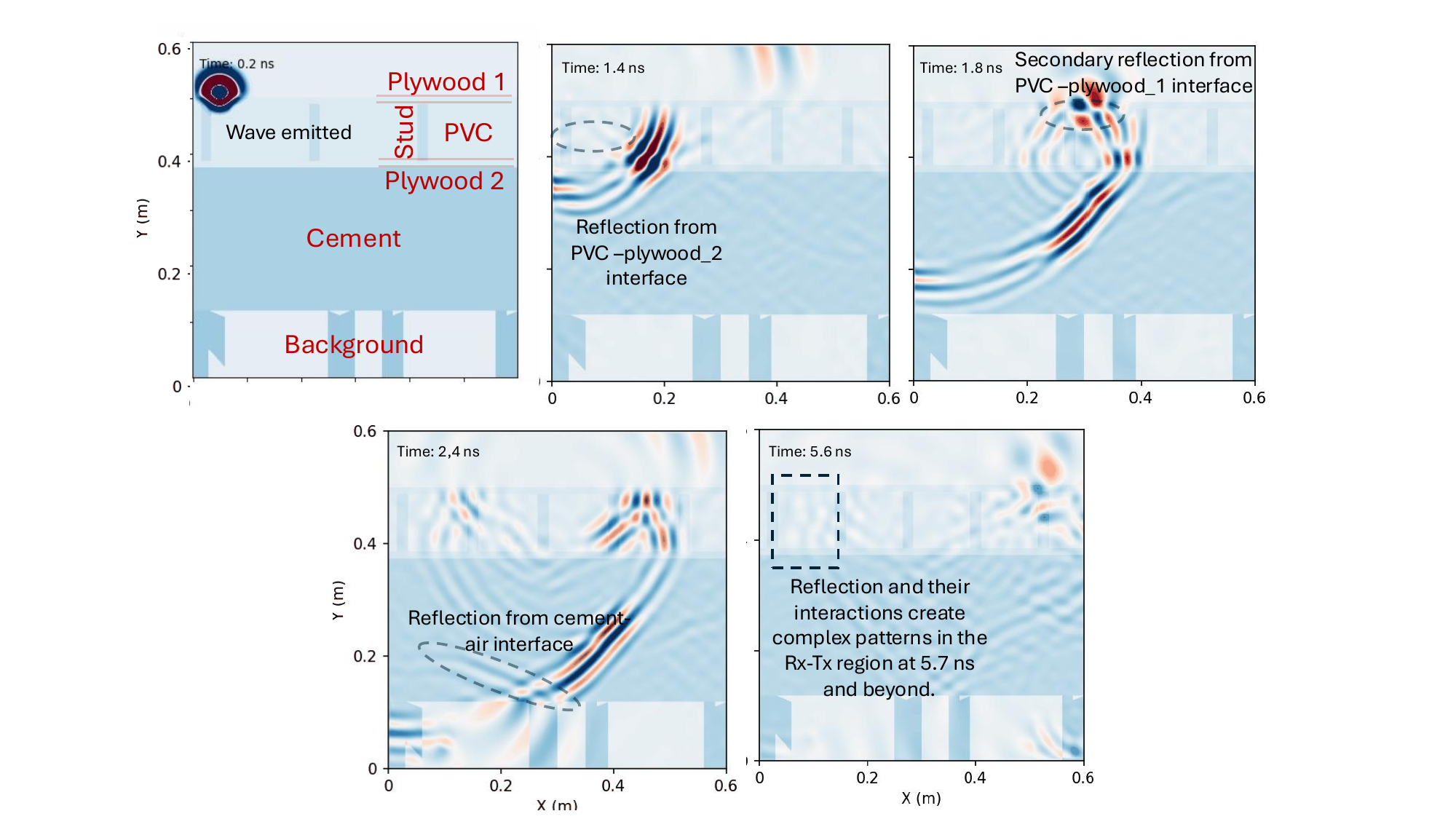}
    \caption{The different structural elements within the envelope produce overlapping reflections that interact to form complex signal patterns in the received SFCW response. The simulated time-domain signal, reconstructed using gprMax, illustrates how these features shape the overall response, though it does not represent true wavefront propagation. Instead, it reflects the cumulative arrival of frequency-dependent components reconstructed from the stepped-frequency synthesis.}
    \label{fig:envelope_animation}
\end{figure}

Compared to conventional GPR applications (e.g., rebar detection in concrete or geological surveys), building envelopes exhibit denser and more systematic structural variations. These variations lead to closely spaced perturbations, as illustrated in Figure~\ref{fig:envelope_animation}. The resulting reflections are often too proximate to be resolved as separate peaks in the A-scan, complicating the attribution of signal features to individual components. A simulated time-domain response, reconstructed from SFCW data using a customized application of the open-source FDTD solver \texttt{gprMax}\cite{GPRMax}, highlights how these structural features shape the received signal over time.

\subsection{Methods}
\label{sec:methods}

The inversion of GPR scans for building envelopes in this study was framed as a set of classification tasks. In future work, this framework can be generalized to other application scenarios where a relevant design for the structure being diagnosed is available.

The classification tasks are as follows: 

\begin{enumerate}

    \item Classify A-scans as stud or non-stud to identify the stud locations. Studs are the only lateral variations that occur within the walls. 
    \item Classify A-scans into different categories that represent the envelope cross-section configuration. There are two such categories for this study.
    \begin{enumerate}
        \item `Interior' for interior walls that are not exposed to the exterior on any side.
        \item `Exterior' for exterior walls with/without background object on either side
    \end{enumerate}
\end{enumerate}

\begin{table}[H]
\centering
\caption{Distribution of wall types and stud labels in the dataset.}
\label{tab:dataset_summary}
\begin{tabular}{llllr}
\hline
\textbf{Label type} & \textbf{Class}   & \textbf{Count} & \textbf{Percentage} & \textbf{Scans} \\
\hline
Wall type & Interior     & 13251 & 44.7\% & \texttt{D1}, \texttt{D2}, \texttt{D3}, 
\texttt{E1}, \texttt{E2}, \texttt{E3}, 
\texttt{F1}, 
\texttt{G1}, \texttt{G2}, \texttt{G3}
 \\
          & Exterior     & 16414 & 55.3\% & \texttt{A1}, \texttt{B1}, \texttt{B2}, \texttt{B3}, 
            \texttt{C1}, \texttt{C2}, \texttt{C3}, 
            \texttt{H1}, \texttt{H2}, \texttt{H3}, 
            \texttt{I1}
             \\
\hline
Stud (SVD) & Stud        & 26721 & 90\% & -- \\
           & Non-stud    & 2944 & 10\% & -- \\
\hline
\end{tabular}
\end{table}

\begin{figure}[H]
    \centering
    \includegraphics[width = 1\linewidth]{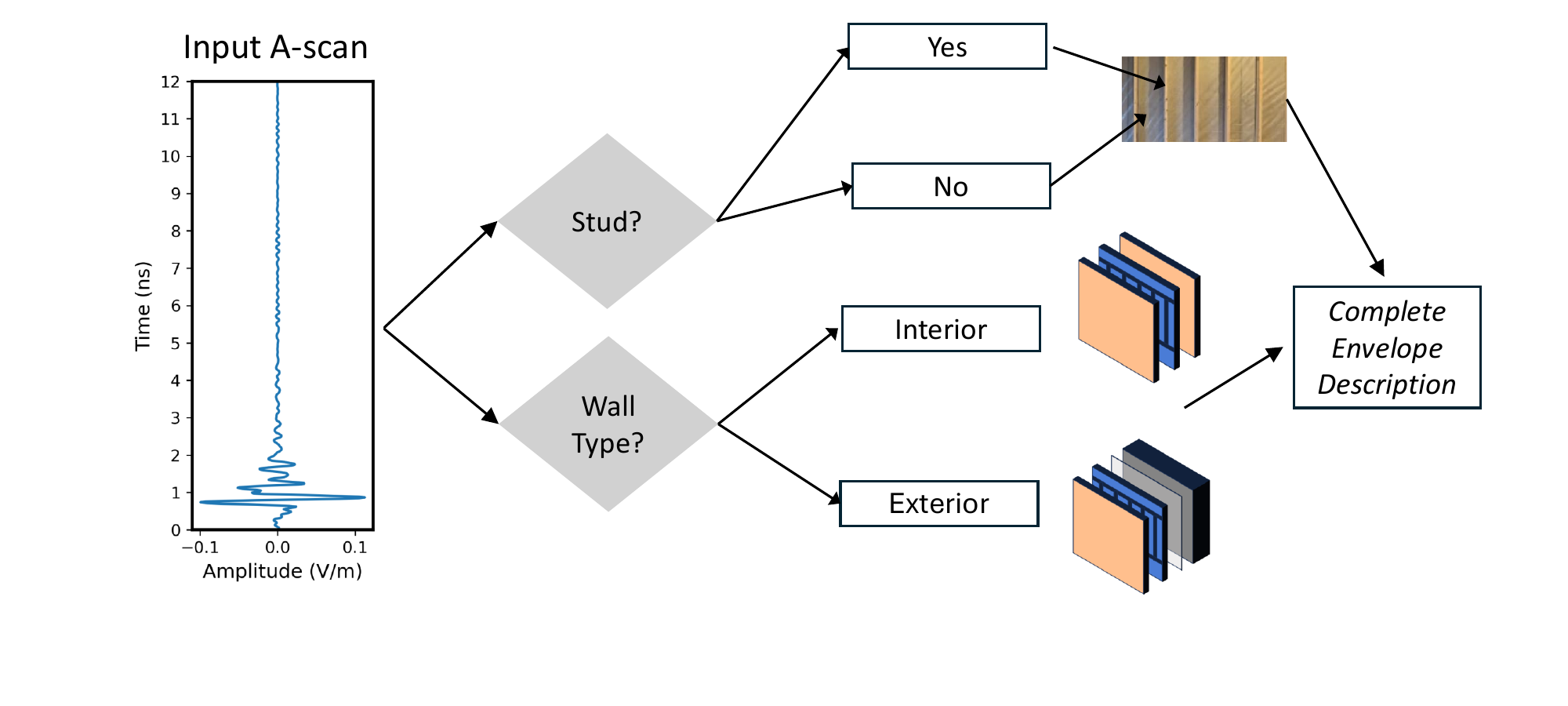}
    \vspace{-1cm}
    \caption{The inversion process involves identifying the cross-sectional variations by identifying the wall type and the lateral variations by detecting the studs.}
    \label{fig:methodology_flow_chart}
\end{figure}

The abovementioned classification tasks provide adequate information to completely describe a building envelope, effectively inverting the GPR scan.

For the stud detection task, Singular Value Decomposition (SVD) is used as a ground-truth indicator. SVD is a matrix factorization technique that decomposes a data matrix into orthogonal components that capture its dominant patterns. In this study, each B-scan (composed of A-scans as columns) is decomposed using SVD to identify recurring structural patterns. The first principal component captures the most variance and reliably tracks periodic perturbations caused by studs, enabling unsupervised identification of their locations.

Four widely used machine learning classification models - K-Nearest Neighbors (KNN), Decision Trees (DT), Random Forests (RF), and Support Vector Machines (SVM - are used as initial base predictors for the classification tasks in this study. KNN is a non-parametric method that classifies each A-scan by aggregating the labels of its k closest neighbors in the time-domain feature space, where closeness is measured using Euclidean distance. In this study, distance-weighted voting is used, giving greater influence to neighbors closer to the query point. DT builds an interpretable tree-like structure by recursively splitting the input features to maximize class separation. At each node, the split is chosen to minimize the Gini impurity - a metric that quantifies the probability of misclassifying a randomly chosen element. RF extends DT by training an ensemble of decision trees on random subsets of both samples and features; predictions are made via majority voting. This ensemble approach reduces overfitting and improves generalization, making it well-suited to the noisy, high-dimensional nature of GPR trace data. SVM classifies data by finding the optimal hyperplane that maximizes the margin between classes; for non-linear separations, it uses kernel functions (e.g., RBF, polynomial) to project data into higher-dimensional spaces where separation is possible, with the decision boundary defined by a subset of training points called support vectors.

To identify the most informative time-domain features in each A-scan and reduce input dimensionality, several feature analysis techniques were applied. Feature agglomeration uses hierarchical clustering to group highly correlated features, either by Euclidean or cosine similarity, and replaces each group with its mean (pooled) or a representative (exemplar) feature. Permutation Feature Importance (PFI) evaluates the effect of each feature on model performance by measuring accuracy drops when its values are randomly permuted - highlighting features that directly contribute to prediction. Finally, Recursive Feature Elimination with Cross-Validation (RFECV) iteratively removes the least important features (based on model-derived importance scores), retraining the model at each step, and selects a minimal feature subset that yields peak cross-validated accuracy. For brevity, the results from these methods are discussed in general terms in the main paper, with detailed analysis and supporting figures provided in the Appendix.

A core contribution of this work is the integration of Sparse Neural Networks (SparseNNs) for GPR signal interpretation. These are neural models that prune uninformative connections dynamically during training, leaving only a subset of active weights at convergence. Unlike external feature selection methods, SparseNNs embed sparsity directly into the learning process, enabling the model to focus on the most informative regions of the time-domain signal while reducing model complexity. By selectively eliminating low-importance connections, SparseNNs perform feature pruning, making the model’s decision process more interpretable and computationally efficient while retaining much of the expressive power of dense architectures. In this work, sparsity is enforced through \textbf{$\mathbf{L_0}$ regularization}, wherein individual weights are stochastically gated and can be driven exactly to zero, effectively pruning connections during training. In contrast, \textbf{$\mathbf{L_1}$ (Lasso)} and \textbf{$\mathbf{L_2}$ (Ridge)} regularization shrink weights continuously toward zero but rarely eliminate them completely. Achieving true sparsity thus requires manual pruning and retraining, which introduces instability, threshold sensitivity, and a dense-to-sparse mismatch. ${L_0}$ regularization instead yields hard sparsity natively, producing compact networks in a single training cycle without post hoc adjustments. Although less commonly used in the past due to its non-differentiable nature and implementation complexity, the \textit{hard-concrete relaxation} formulation~\cite{hard_concrete_relaxation} enables gradient-based optimization through differentiable stochastic gates that approximate binary dropout. The resulting networks remain differentiable during training yet contain exact zeros at convergence, offering both computational efficiency and physical interpretability. These properties—efficiency, compactness, and transparency—make SparseNNs powerful analytical tools for GPR interpretation and potentially practical for edge-computing scenarios, where limited resources demand lightweight yet accurate models.

To assess the significance of features selected by the SparseNN models, we applied SHAP (SHapley Additive exPlanations)\cite{lundberg}, a model-agnostic interpretability framework grounded in cooperative game theory. SHAP assigns each input feature a Shapley value by comparing predictions with and without that feature across all possible subsets of the remaining features. Because models cannot generally accept inputs with features literally removed, SHAP simulates “missingness” by replacing features with typical values drawn from a background distribution (here, representative test scans). This ensures that each importance value reflects the marginal contribution of a feature relative to a baseline expectation—the average model output with no feature information. The SparseNN first identified a trimmed feature set by selecting a subset of time-domain indices during training. SHAP analysis was then applied to this reduced set of features to quantify each feature's contribution to model predictions, thereby assessing how the most salient reflections influenced classification. For each selected feature, SHAP values were obtained across the evaluation set, though only a representative subset of samples (on the order of hundreds) was visualized to illustrate distributional trends. By providing a consistent, locally accurate, and model-agnostic measure of feature importance, SHAP serves as a post hoc diagnostic tool for examining the alignment between learned sparsity and model behavior, offering a complementary perspective on feature relevance that builds on broader interpretability practices in machine learning\cite{molnar_interpret}.

To further assess the physical significance of selected features, wave propagation calculations were performed to map signal arrival times to plausible reflection points within the wall geometry. By using representative permittivity values for typical building materials (e.g., drywall, insulation, cement), approximate wave travel paths and depths were estimated for each feature’s time location (the method is further detailed in Appendix: Figure~\ref{fig:wave_propagation_calculations}). This analysis enables spatial interpretation of the selected features, linking specific time-domain signals to structural features such as stud faces, material interfaces, and layer boundaries. By comparing these estimated reflection paths to known wall layouts, the study evaluates whether the features selected by SparseNNs or other models correspond to physically meaningful locations in the envelope, thus reinforcing interpretability from a physics-based perspective.

\section{Results}

\subsection{Stud Detection}

Studs create the most prominent perturbation in the B-scans present in the dataset. A typical approach to developing stud-detection models would have been to create a structural map of stud distributions and subsequently train ML models to classify A-scans as stud/non-stud. However, such a method would require significant manual effort, requiring precise measurements of multiple building walls. It would also involve tearing down the wall envelope in certain regions. 

\begin{figure}[H]
    \centering
    \includegraphics{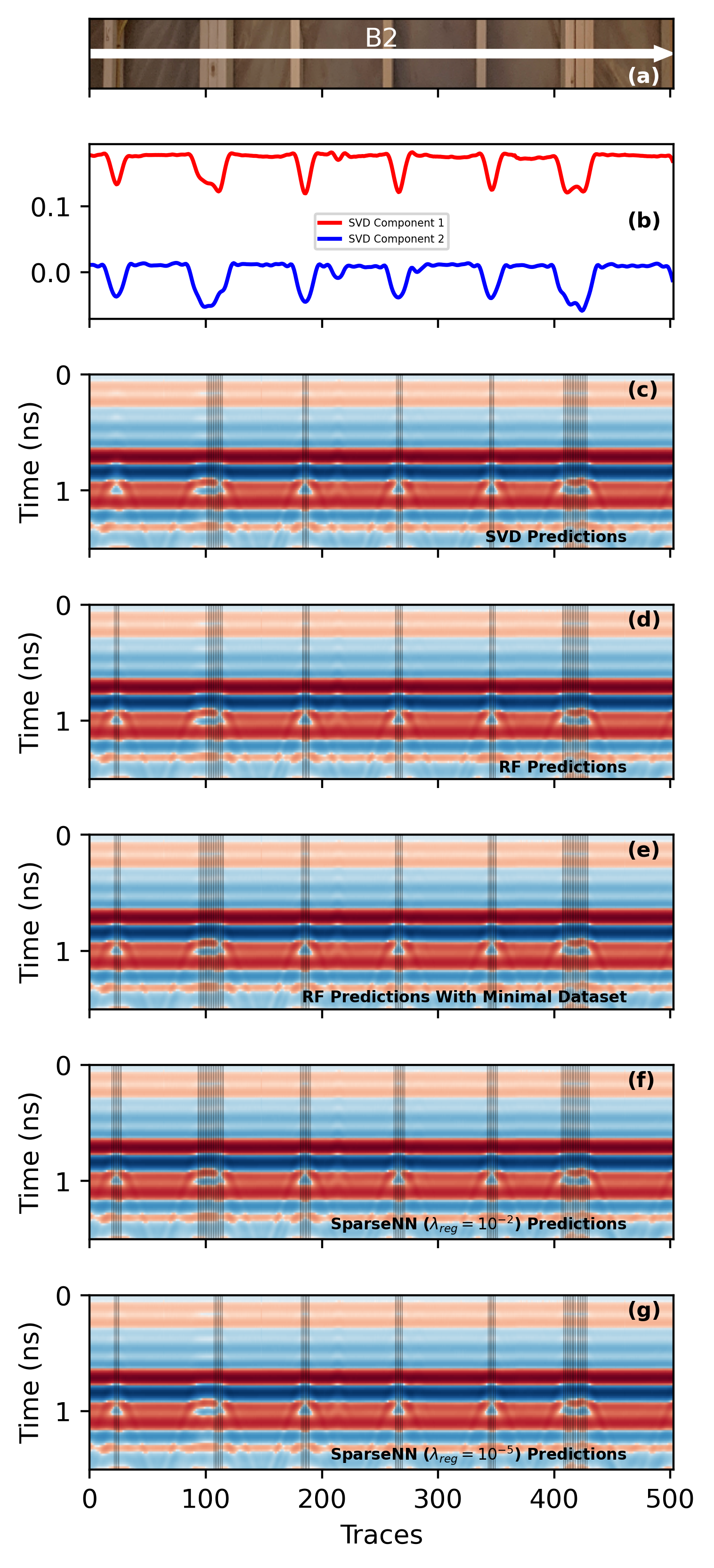}
    \caption{Vertical lines indicate predicted stud locations.
    (a) Wall interior corresponding to Scan \texttt{B2}; 
    (b) principal SVD components fluctuations align closely with stud locations; 
    (c) SVD-based stud predictions; 
    (d) Stud predictions using RF, where the RF model detects the left-most stud missed by the SVD method. 
    Subsequent variants of RF and SparseNN (panels e--g) also detect the stud missed by SVD.}
    \label{fig:stud_detection_predictions}
\end{figure}

An alternative to this invasive and cumbersome sample-collection approach was to use prior knowledge of stud dimensions to automatically generate accurate sample labels. The studs in this residential building have a nominal surfaced thickness of 1.50~in, but sawmills typically produce slightly oversized lumber—ranging from 1.51 to 1.54~in~\cite{ALSC_NGR2017}. After installation, the equilibrium moisture content decreases and shrinkage of approximately 0.03~in or more is common~\cite{FPL_WoodHandbook2021}. These figures together indicate an expected dimensional tolerance of at least 0.03~in for stud width.

To utilize this knowledge, Singular Value Decomposition (SVD) (see Section:~\ref {sec:methods}), was applied to each B-scan (with each A-scan as an individual data sample). It was found that both the primary and secondary components of SVD tracked strongly with the stud locations in the envelope (Figure~\ref{fig:stud_detection_predictions}b). 

Initial stud labeling was performed on a scan-by-scan basis using the first principal component of the SVD for each B-scan. As shown in Figure~\ref{fig:stud_detection_predictions}(b), studs correspond to the extreme points (either maxima or minima) of the SVD component across the scan. For each B-scan, the three most prominent outliers in the distribution of the first SVD component were identified. If these outliers exceeded the component’s mean, the local maxima were labeled as studs; conversely, if they were below the mean, the local minima were labeled as studs. A threshold fraction (Appendix:~Figure~\ref{fig:stud_detection_predictions}) was then used to delineate the extent of each peak, defining where a stud begins and ends.

Prior knowledge of stud thickness was subsequently incorporated into the detection model by tuning the threshold fraction—defined as a fraction of the SVD component's mean—so that the detected studs most frequently exhibited a width of 1.50~in (3.81~cm). This calibrated threshold fraction also yielded an overall mean stud width close to 1.50~in (Appendix:~Figures~\ref{fig:stud_detection_svd_cut_off},~\ref{fig:stud_width-distribution}).

As seen in Figure~\ref{fig:stud_detection_predictions}(c), the predictions derived from the SVD component perturbations align closely with the actual stud locations in the envelope, although the leftmost stud remains undetected. However, because this method requires prior knowledge of the envelope—specifically the true stud widths—it becomes inapplicable in situations where such information is unavailable.

To develop a more general framework, the stud locations predicted from the SVD component oscillations were used as ground truth for training classical machine learning (ML) classification models. Training data were selected from scans where the predicted studs exhibited more uniform widths, as these cases more accurately reflected the true geometry. Predictions showing stacked studs that appeared noticeably wider than single studs were also preferred for the same reason. Scans \texttt{I1} and \texttt{A1} (interior walls) and \texttt{G1}-\texttt{G3} (exterior walls), which satisfied these criteria, were chosen for training.

Four widely used machine learning classifiers - Decision Trees (DT), Random Forest (RF), K-Nearest Neighbors (KNN), and Support Vector Machines (SVM) - were applied to the stud detection problem (see Sec.~\ref{sec:methods}). Among these, RF achieved the highest performance, with perfect (100\%) accuracy on the training set and $0.985 \pm 0.001$ accuracy on the held-out test set.

In several cases, the RF model outperformed the SVD-based approach by generalizing better to new scans, as illustrated in Figure~\ref{fig:stud_detection_predictions}(c). In scan \texttt{B2}, the leftmost stud missed by the SVD method was correctly identified by the RF model, although its width appeared slightly underestimated.

While these results are encouraging, they do not by themselves establish the method’s practical feasibility or generalizability. The framework relied on large, contiguous wall sections as training data, making it inherently invasive and destructive. Moreover, the opaque nature of the ML models limited interpretability; it remained unclear which features contributed to each prediction or how they influenced the decision-making process. As shown earlier in Figure~\ref{fig:envelope_animation}, the various envelope components produce complex scattering patterns; the models may therefore be learning from far-field and/or higher-order reflections within the wall, which would restrict their broader applicability. Further analysis was thus required to evaluate the robustness and generalizability of the approach across diverse conditions.

\subsubsection{Information Minimization}

To ensure practical applicability in real-world settings where labeled data may be limited and data acquisition costly, this study investigates model performance under minimal information conditions. Minimization is considered along two dimensions: (1) sample minimization, where only a small subset of the data is used for training, and (2) feature minimization, where only a sparse subset of the 655 time-sample features per trace is used. Importantly, data sparsity can sometimes enhance model fidelity by focusing learning on the most informative features while discarding spurious or redundant inputs. This aligns with established machine learning theory, which holds that simpler, more parsimonious models tend to generalize better and offer greater interpretability~\cite{guyon2003info_minimization}.

\paragraph{Sample Minimization}\leavevmode\newline

In field applications, accurately labeled data are scarce. Therefore, the proposed stud detector is evaluated using a minimal dataset: only one interior scan (\texttt{I1}) and one exterior scan (\texttt{G3}), corresponding to two wall segments within the house.

\begin{table}[H]
\centering
\begin{tabular}{ c c }
\hline
\textbf{Train} & \textbf{Test} \\
\hline
\texttt{G3 and I1} & All other scans \\
\hline
\end{tabular}
\caption{Minimal train set and test set for stud detection}  
\label{tab:train_test_minimum_stud_detection}
\end{table}

Using the minimal training set, the default RF model achieved an accuracy of $0.964 \pm 0.002$ on the test set.

\subsubsection{Feature Minimization}

To evaluate and improve the efficiency of stud detection, several feature minimization strategies were explored and are detailed in Appendix: Section~\ref{sec:appendix_stud_detection_feature_minimization}. Feature agglomeration using hierarchical clustering demonstrated that only 3–10 pooled features (via Euclidean distance) could retain performance close to that of the full 655-feature model. In contrast, exemplar-based agglomeration, which avoids pooling, yielded lower but interpretable accuracy, establishing a conservative performance baseline. Permutation Feature Importance (PFI) analysis confirmed that only a narrow time window (1 ns) carried significant predictive value, while most features beyond 2 ns contributed negligible or even negative importance. Recursive Feature Elimination with Cross-Validation (RFECV) further supported this sparsity, achieving $0.966 \pm 0.001$ test accuracy with just 95 features using standard stratified folds, and $0.981 \pm 0.001$ with group-based folds—despite variability introduced by scan-level partitioning. These results collectively highlight that the majority of the A-scan signal is redundant or noisy for this task, and that accurate stud detection is achievable using highly compact, physically interpretable feature subsets.

\indent\subparagraph{Sparse Neural Network}\leavevmode\newline

Building on these findings, we next investigated whether sparsity could be enforced directly within the learning architecture, which motivated the use of Sparse Neural Networks (SparseNNs; see Section~\ref{sec:methods}).
Here, sparsity is applied during training via a regularization term in the loss function, controlled by a hyperparameter ($\lambda_{\text{reg}}$), to balance accuracy with structural simplicity. This enables the model to discard redundant connections while retaining, or even improving, generalization.

A single-layer SparseNN with only eight hidden neurons was trained for stud prediction across varying values of $\lambda_{\text{reg}}$, using the minimal dataset described in Table~\ref{tab:train_test_minimum_stud_detection}. As noted earlier, studs produce the most prominent perturbations in the scans, which likely explains why increasing the number of layers or neurons did not yield meaningful performance gains; the task is too simple to benefit from additional depth or capacity.

As shown in Table~\ref{tab:stud_detection_SparseNN_performance_summary} (and Appendix: Figure~\ref{fig:stud_SparseNN_performance}), even when restricted to a single input feature (green line), the compact SparseNN achieves performance comparable to the Random Forest (RF) model using all available features (approximately 0.935 accuracy for SparseNN vs. 0.985 for RF). As the regularization strength decreases, both model performance and the number of active features increase. Notably, with 11 input features, the SparseNN surpasses all RF variants in accuracy, demonstrating the effectiveness of sparse training in identifying compact yet highly predictive feature subsets.

\begin{table}[H]
\centering
\begin{tabular}{ c c c }
\hline
\textbf{Model + Feature Elimination Process } & \textbf{Train/Validation Accuracy} & \textbf{Test Accuracy} \\
\hline
RF (all features, comprehensive dataset) & 1.0 $\pm$ 0.0 & 0.985 $\pm$ 0.001 \\
\hline
RF (all features, minimal dataset) & 1.0 $\pm$ 0.0 & 0.964 $\pm$ 0.002 \\
\hline
RF (RFECV - Stratified) & 0.998 $\pm$ 0.003 & 0.966 $\pm$ 0.001\\
\hline
RF (RFECV - Stratified Grouped) & 0.984 $\pm$ 0.002 & 0.981 $\pm$ 0.001 \\
\hline
\hline
RF (Euclidean Agglomeration) & 1.0 $\pm$ 0.0  & 0.963 $\pm$ 0.001 \\
\hline
RF (Cosine Agglomeration) & 1.0 $\pm$ 0.0  & 0.880 $\pm$ 0.001 \\
\hline
RF (Exemplar Agglomeration) & 1.0 $\pm$ 0.0  & 0.934 $\pm$ 0.001 \\
\hline
\hline
\textbf{SparseNN (minimal dataset)} & \textbf{1.0 $\pm$ 0.0} & \textbf{0.983 $\pm$ 0.004}\\
\hline
\end{tabular}
\caption{Summary of stud detection performances with different models and feature selection criteria.}  
\label{tab:stud_detection_performance_summary}
\end{table}

\begin{table}[H]
\centering
\begin{tabular}{ c c c c c }
\hline
\textbf{Architecture} & \textbf{$\lambda_{reg}$} & \textbf{$n_{features}$} & \textbf{Train Accuracy} & \textbf{Test Accuracy} \\
\hline
(8,) - \textbf{fails to converge} & $10^{-1}$ & 593 & 0.583 $\pm$ 0.404 & 0.580 $\pm$ 0.391\\
(8,) & $10^{-2}$ & 1 & 0.981 $\pm$ 0.023 & 0.935 $\pm$ 0.012 \\
(8,) & $10^{-3}$ & 2 & 1.000 $\pm$ 0.000 & 0.955 $\pm$ 0.004 \\
(8,) & $10^{-4}$ & 8 & 1.000 $\pm$ 0.000 & 0.977 $\pm$ 0.006 \\
\textbf{(8,)} & \textbf{$10^{-5}$} & \textbf{11} & \textbf{1.000 $\pm$ 0.000} & \textbf{0.983 $\pm$ 0.004} \\
\hline
\end{tabular}
\caption{Stud detection performance across regularization strengths. Peak performance occurs with $\lambda_{reg} = 10^{-5}$, utilizing 11 features.}
\label{tab:stud_detection_SparseNN_performance_summary}
\end{table}




\subsubsection{Feature Robustness and Locations}

To evaluate the robustness of feature selection in the SparseNN model for the stud detection task, 20 models were trained using the same train–test split and the optimized hyperparameters corresponding to the most feature-efficient configuration in Table~\ref{tab:stud_detection_performance_summary} ($\lambda_{\text{reg}} = 10^{-2}$). Because SparseNN performs stochastic feature pruning, repeated convergence to the same features across multiple runs would indicate that those features are physically meaningful, reflecting stable signal patterns linked to material interfaces or studs. In contrast, features arising from noise or spurious correlations are unlikely to persist across independent trainings.

\begin{figure}[H]
    \centering
    \includegraphics[width = 0.8\textwidth]{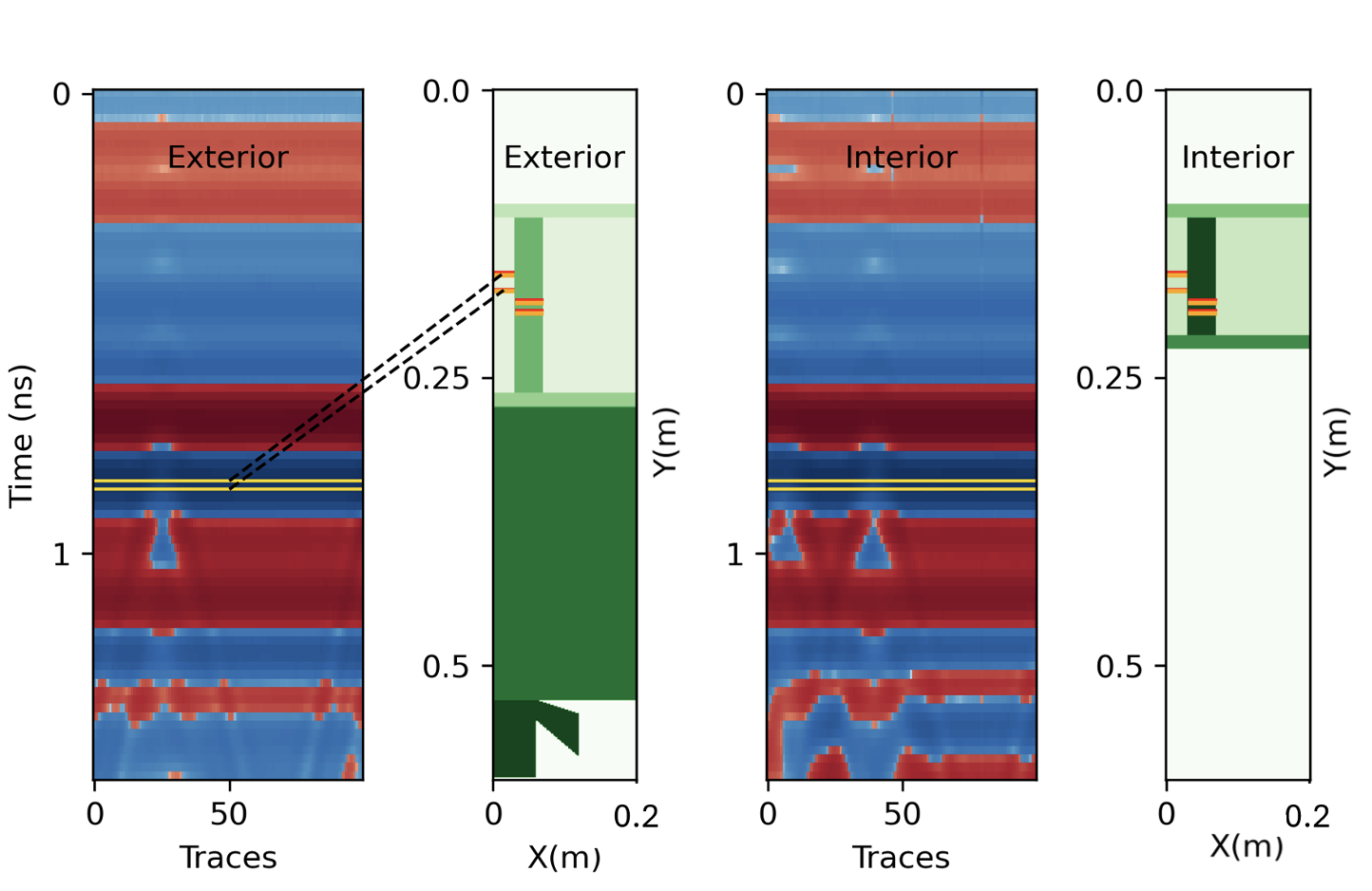}
    \caption{With $\lambda_{\text{reg}} = 10^{-2}$, repeated SparseNN runs consistently converged on one of two signal samples (0.843~ns or 0.861~ns), shown as gold lines on the B-scans. Wave-propagation analysis (Appendix: Section~\ref{sec:wave_propagation_calculation}), using representative permittivity values for common building materials, confirmed that both features originated from regions of structural variation associated with studs. Each color in the geometry map corresponds to one of the two selected features (time samples), with the paired lines of the same color (orange or yellow) representing wave paths computed using the upper and lower bounds of the assumed material permittivity range.}
    \label{fig:stud_feature_locations_robustness}
\end{figure}

 Repeated training runs of the SparseNN model consistently converged on one of two specific features in the signals, corresponding to time samples at 0.843~ns and 0.861~ns (marked by gold lines on the B-scans). Using representative permittivity values for common building materials, wave-propagation analysis (Appendix: Section~\ref{sec:wave_propagation_calculation}) indicated that these reflections originated from regions within the wall envelope where structural variations due to studs were present. Each color in the geometry map corresponds to one of the two selected features (time samples), with the paired lines of the same color (orange or yellow) representing wave paths computed using the upper and lower bounds of the assumed permittivity range.

In summary, the results demonstrate that accurate stud detection can be achieved using minimal data and compact, interpretable models. SparseNNs not only matched the performance of traditional ensemble methods but also revealed physically consistent features that align with known wall geometry. The convergence of feature selection near stud-induced perturbations, supported by wave-propagation analysis, highlights the model’s ability to learn representations grounded in the underlying structural physics rather than data correlations.

\subsection{Wall Classification}
\label{sec:wall_classification}

The second classification task involved the categorization of wall and background combinations. Although the differentiation between interior and exterior walls is straightforward regarding data labeling, that is not the case with wall backgrounds. Many exterior wall segments have gradual transitions in the background, making it difficult to determine a cut-off point (Figure~\ref{fig:wall_variations}). 

\begin{figure}[H]
    \centering
    \includegraphics{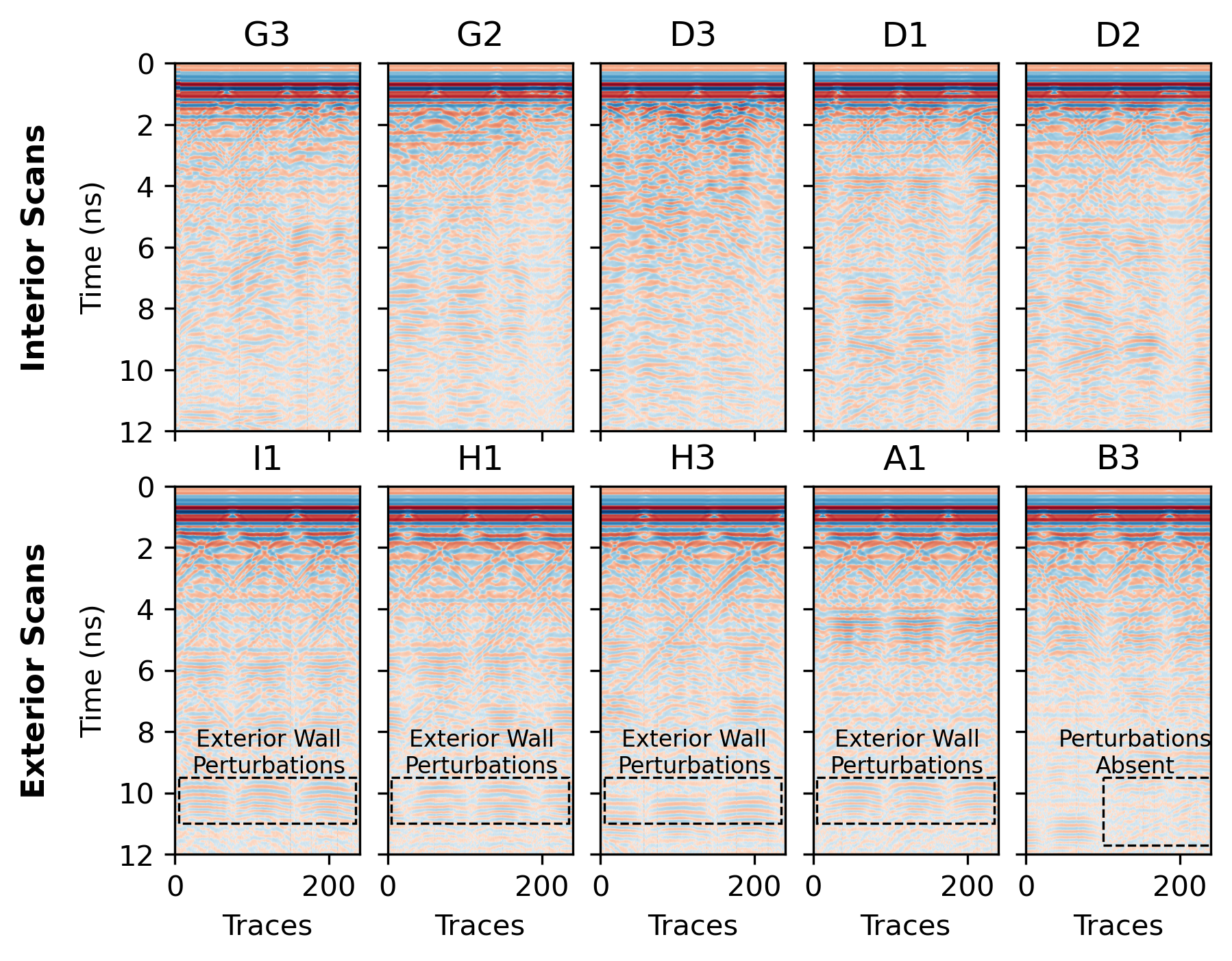}
    \caption{Comparison of exterior and interior wall segment scans. Each wall segment produces distinct signal patterns. Although some patterns do seem to persist between wall types (for instance, the regular reflections around 10 ns in the external wall scans), they are not consistent (the signals are shifted and absent in the rightmost traces of exterior scan \texttt{B3}).}
    \label{fig:wall_class_data_variations}
\end{figure}

Moreover, despite the significant difference between interior and exterior envelopes in composition layers, the two wall types did not have distinct signal patterns that separated them. Each wall segment produces its unique signal patterns. Although some patterns do seem to persist between wall types (for instance, the regular reflections at around 10~ns in the external wall scans), they are not consistent (the signals are quite shifted at the leftmost traces and absent in the rightmost traces of external scan \texttt{B3} in Figure~\ref{fig:wall_class_data_variations}).

Similar to the stud detection task, Decision Trees (DT), Random Forests (RF), K-Nearest Neighbors (KNN), and Support Vector Machines (SVM) were initially applied to the wall and background classification problem, using the comprehensive train and test sets described in Table~\ref{tab:train_test_set_wall_classification}. As noted earlier, certain wall regions contain ambiguous background labels due to gradual transitions. To avoid introducing errors from an arbitrary cut-off, the training set was restricted to scans with a single definitive class, while scans exhibiting background transitions (\texttt{H1} and \texttt{H3}) were allocated to the test set.

\begin{table}[!h]
\centering
\begin{tabular}{ c c c c }
\hline
\textbf{Wall + Background Type} & \textbf{Wall Type} & \textbf{Train} & \textbf{Test} \\
\hline
interior & interior & \texttt{G1} & \texttt{G2-G3} , \texttt{E1-E3}, \texttt{D1-D3,F1} \\
\hline
exterior & exterior & \texttt{B1-B2}  & \multirow{2}{*}{\makecell{\texttt{H1,H3} \\ (exterior $\rightarrow$ exterior + earth)}} \\
\cline{1-3}
exterior + earth & exterior & \texttt{I1,H2,A1} &  \\
\hline
\end{tabular}
\caption{Comprehensive training and test sets used for the wall and background classification task.} 
\label{tab:train_test_set_wall_classification}
\end{table}

As in the stud detection task, the Random Forest (RF) model achieved the best performance among the tested classifiers, reaching perfect accuracy on the training set and $0.805 \pm 0.024$ on the test set.
 

However, background classifications with the RF model approached random accuracy, indicating that the models were unable to extract reliable information from these signals. None of the tested ML approaches could successfully differentiate exterior walls with a background (“exterior + earth”) from those without a background (“exterior”). A likely explanation is that the backgrounds introduce highly variable responses, reflecting soil, rock patterns, and other heterogeneous clutter that do not yield consistent features across samples. In addition, these background interfaces are physically more distant from the antenna, so their contributions to the received signal are weaker and often obscured by attenuation and overlapping clutter. Together, these factors make background classification substantially more challenging than tasks focused on near-surface structural features such as studs or wall type. Subsequent experiments therefore focused on pure wall classification—distinguishing between interior and exterior wall types—a task that still carries substantial diagnostic value.


\subsubsection{Minimization}
\paragraph{Sample Minimization}\leavevmode\newline

As in the stud detection task, a minimal dataset, requiring minimal invasive hypothetical data collection, was used to train the RF model. Only scans \texttt{I1} and \texttt{G3}, representing exterior and interior walls, were included in training. 

Using this minimal dataset, the RF model achieved an accuracy of $0.799 \pm 0.026$ on the test set.

\paragraph{Feature Minimization}\leavevmode\newline

To reduce model complexity and improve interpretability, multiple feature minimization strategies were applied to the wall classification task (described in Appendix: Section~\ref{sec:appendix_wall_classification_feature_minimization}). Agglomeration-based methods  revealed that clustering temporally correlated A-scan features using cosine similarity significantly boosted Random Forest (RF) performance, with only 15 clusters achieving near-peak accuracy. In contrast, exemplar-based agglomeration (which retains a single feature per cluster) performed substantially worse, suggesting that pooling was crucial for this task. Permutation Feature Importance showed that important features for wall classification were more distributed across the signal than in stud detection, extending up to 6 ns, indicating contributions from higher-order reflections relevant for this task. Recursive Feature Elimination with Cross-Validation further refined the input space, achieving 0.850 $\pm$ 0.003 test accuracy with only 95 features using a stratified group CV setup. 

\indent\subparagraph{Sparse Neural Network}\leavevmode\newline

Similar to the stud detection task, the coexistence of a few highly informative features with many spurious ones motivated the use of SparseNNs for wall classification.

Unlike in stud detection, where only a single SparseNN architecture was examined, multiple architectures of varying depth were explored here. This broader search was motivated by the expectation that features relevant to wall type are both more numerous and more widely distributed across the A-scan. Increasing network width or depth can capture higher-order abstractions of these signals, potentially improving performance while, somewhat counterintuitively, yielding even sparser and more selective feature representations.

To account for this possible variation, we trained a range of SparseNN architectures under varying regularization strengths. Appendix: Figure~\ref{fig:wall_classification_SparseNN_performance} illustrates how representative models evolved during training, highlighting both classification accuracy and progressive feature pruning. Among the regularization strengths tested, $\lambda_{reg} = 10^{-4}$ proved most stable and broadly effective, consistently producing models that converged reliably while maintaining strong sparsity. This value was therefore fixed for subsequent experiments. All SparseNN models were trained on the minimal dataset comprising scans \texttt{I1} and \texttt{G3}, enabling assessment of architectural effects under constrained data conditions.

\begin{table}[H]
\centering
\begin{tabular}{ c c c c }
\hline
\textbf{Architecture} & \textbf{$n_{features}$} & \textbf{Train Accuracy} & \textbf{Test Accuracy} \\
\hline
(16, n) & 8 & 1.000 $\pm$ 0.000 & 0.893 $\pm$ 0.003 \\
(16, y) & 8 & 1.000 $\pm$ 0.000 & 0.898 $\pm$ 0.001 \\
(32, n) & 7 & 1.000 $\pm$ 0.000 & 0.906 $\pm$ 0.004 \\
(32, y) & 9 & 1.000 $\pm$ 0.000 & 0.899 $\pm$ 0.003 \\
(64, n) & 8 & 0.999 $\pm$ 0.002 & 0.905 $\pm$ 0.007 \\
\textbf{(64, y)} & \textbf{9} & \textbf{1.000 $\pm$ 0.000} & \textbf{0.908 $\pm$ 0.007} \\
(8,8, n) & 8 & 1.000 $\pm$ 0.000 & 0.900 $\pm$ 0.001 \\
(8,8, y) & 8 & 1.000 $\pm$ 0.000 & 0.901 $\pm$ 0.001 \\
(16,8, n) & 8 & 1.000 $\pm$ 0.000 & 0.900 $\pm$ 0.001 \\
(16,8, y) & 8 & 1.000 $\pm$ 0.000 & 0.899 $\pm$ 0.001 \\
(32,16, n) & 7 & 1.000 $\pm$ 0.000 & 0.900 $\pm$ 0.001 \\
(32,16, y) & 8 & 1.000 $\pm$ 0.000 & 0.900 $\pm$ 0.002 \\
(8,8,8, n) & 7 & 1.000 $\pm$ 0.000 & 0.898 $\pm$ 0.000 \\
(8,8,8, y) & 9 & 1.000 $\pm$ 0.000 & 0.899 $\pm$ 0.001 \\
(32,16,8, n) & 8 & 1.000 $\pm$ 0.000 & 0.900 $\pm$ 0.002 \\
(32,16,8, y) & 7 & 1.000 $\pm$ 0.000 & 0.907 $\pm$ 0.006 \\
\hline
\end{tabular}
\caption{SparseNN wall-type classification performance across network architectures and stud configurations. All results use a fixed $\lambda_{\text{reg}} = 10^{-4}$ and represent averages over five independent runs. The number of selected features ($n_{\text{features}}$) is also averaged across runs and rounded to the nearest whole number.}
\label{tab:wall_classification_SparseNN_performance}
\end{table}

\begin{figure}
    \centering
    \includegraphics{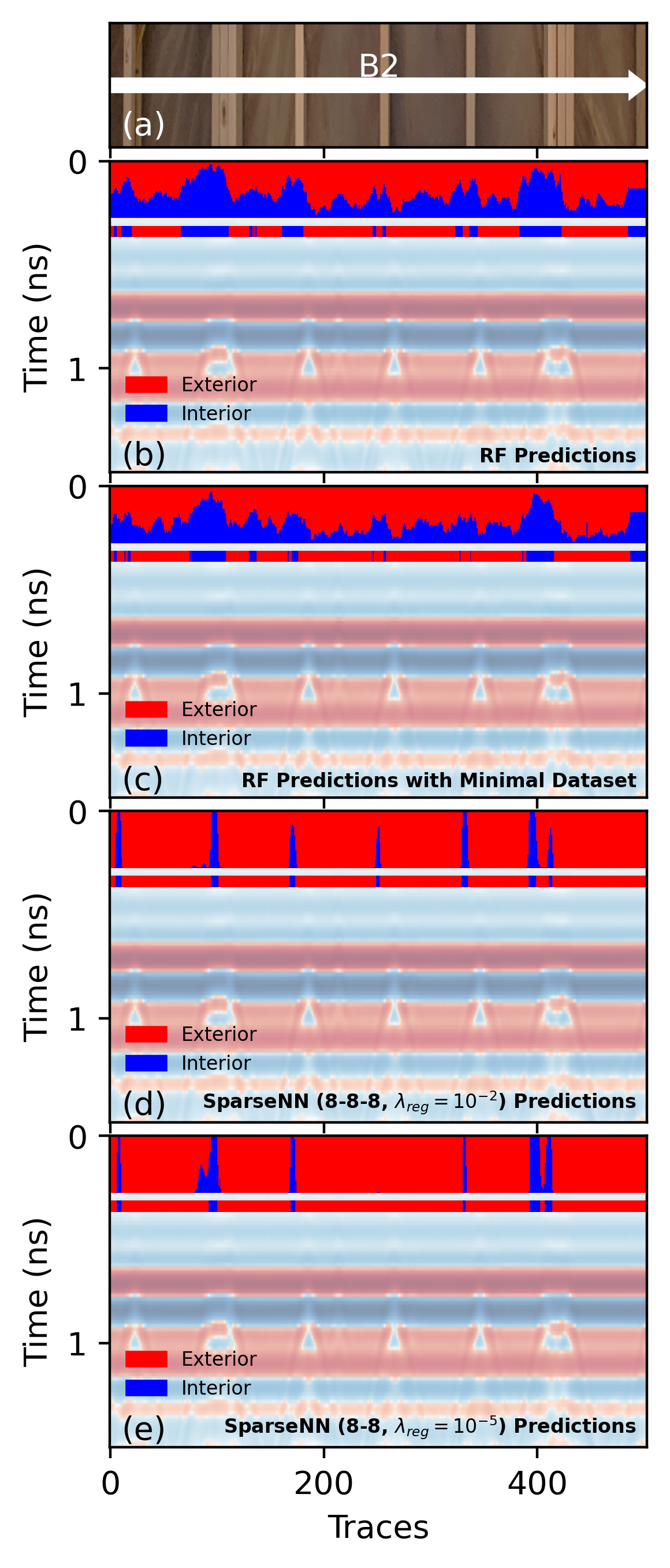}
    \caption{Predictions of various models on the test scan \texttt{B2}. Panel (a) displays an image of the scanned scan, while each panel below follows a common format: the wider band represents predicted class probabilities, and the narrower band below it indicates the final predicted class. All models exhibit misclassifications near stud regions. However, compared to the Random Forest (RF) baseline, the SparseNN models are less affected by the stud perturbations, showing sharper, more confident probability bands and narrower zones of misclassification. All SparseNNs are trained with the minimal dataset.}
    \label{fig:wall_classification_predictions}
\end{figure}

Figure~\ref{fig:wall_classification_predictions} compares the predictions of various models on the test scan \texttt{B2}. The top panel displays an image of the scanned wall segment, while the panels below show predicted class probabilities (wide band) and the corresponding final predicted class (narrow band).

Many of the misclassifications occur near stud regions within the wall. These errors arise from strong perturbations caused by lateral variations in stud geometry, which can confuse trace-based classifiers. However, as seen in Figure~\ref{fig:wall_classification_predictions}, compared to the Random Forest (RF) baseline, the SparseNN models are less affected by the stud perturbations, showing sharper, more confident probability bands and narrower zones of misclassification. 

To test whether this robustness could be further improved, we incorporated explicit stud-location information, derived from prior predictions, by appending a binary indicator of stud presence to each A-scan. This structured encoding of physical context allowed evaluation of whether the network could effectively leverage global structural cues.

\begin{figure}
    \centering
    \includegraphics{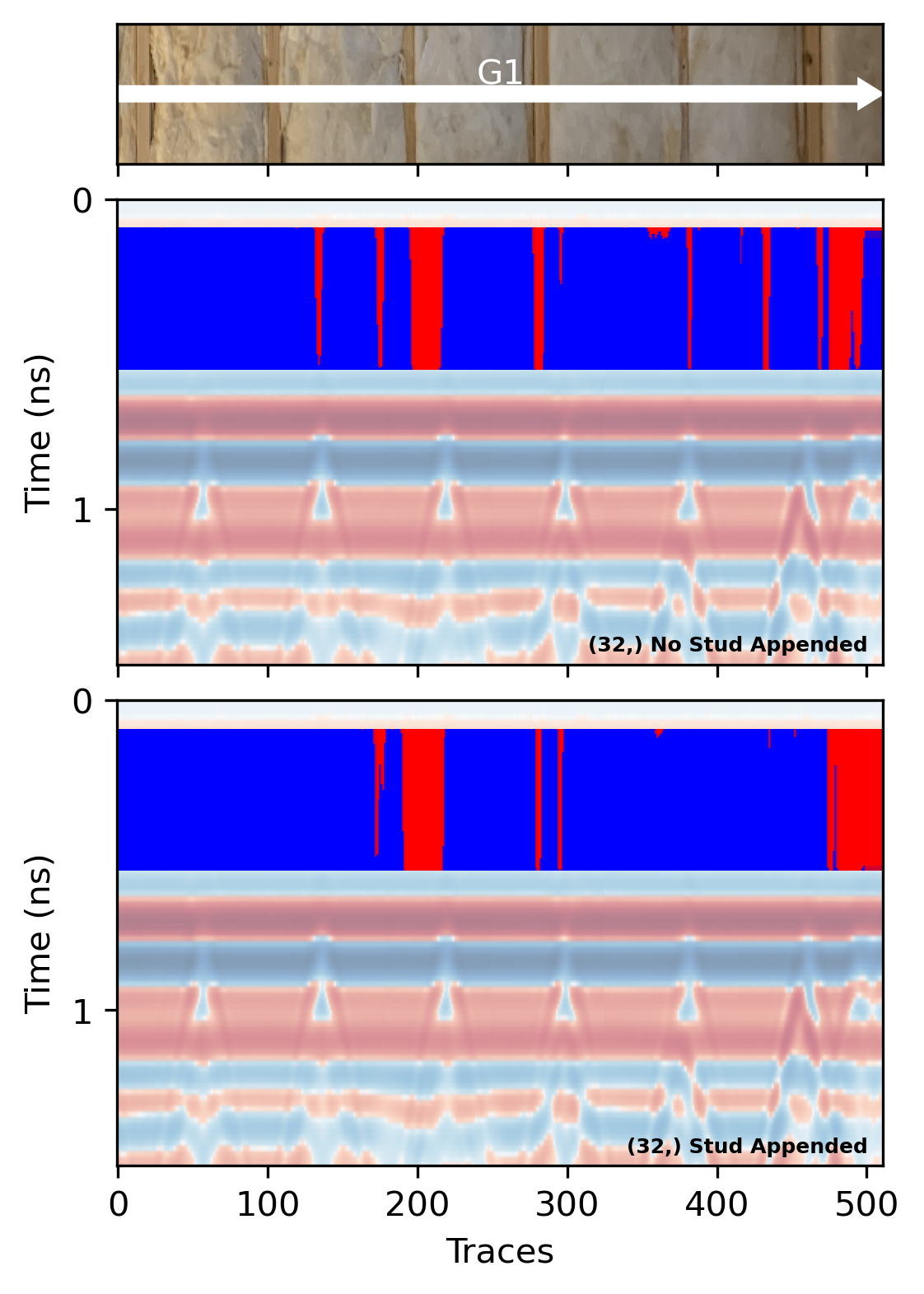}
    \caption{Test set predictions from the same SparseNN architecture (32,) with and without the stud indicator appended to the input. Appending the stud information to the input trace results in fewer misclassifications near stud regions.}
    \label{fig:wall_classification_SparseNN_stud_appended}
\end{figure}

Table~\ref{tab:wall_classification_SparseNN_performance} summarizes these results: in the architecture labels, ‘y’ denotes inclusion of the stud indicator and ‘n’ its absence. The addition of stud context produced only modest gains in overall accuracy but consistently reduced local misclassifications near studs, as illustrated in Figure~\ref{fig:wall_classification_SparseNN_stud_appended}.

Further analysis reveals that the SparseNN models consistently retained the appended stud indicator as part of the final feature set at convergence. This is noteworthy because most informative signal perturbations occur quite early in the 12-ns trace (see Appendix: Figure~\ref{fig:PFI_wall_classification}), well before the appended stud indicator. The network’s selection of this distant, context-encoding feature suggests it is not merely exploiting local correlations but instead learning a higher-level representation of structural context across the wall.

Overall, SparseNNs outperformed all other models and feature elimination combinations (Table~\ref{tab:wall_classifcation_overall_performance_summary}). The best-performing configuration (64, y) achieved a mean accuracy of 0.908, exceeding both the Random Forest baseline (0.799) and the RF with cosine agglomeration (0.899), while utilizing only seven features. Subsequent analysis reveals that these retained features align closely with physically meaningful regions of the signal.




\begin{table}[H]
\centering
\begin{tabular}{ c c c }
\hline
\textbf{Model + Feature Elimination Process } & \textbf{Train/Validation Accuracy} & \textbf{Test Accuracy} \\
\hline
RF (all features, comprehensive dataset) & 1.0 $\pm$  0.0 &  0.805 $\pm$  0.024 \\
\hline
RF (all features, minimal dataset) & 1.0 $\pm$ 0.0 & 0.799 $\pm$ 0.026\\
\hline
\hline
RF (RFECV - Stratified) & 1.0 $\pm$ 0.001 & 0.817 $\pm$ 0.042  \\
\hline
RF (RFECV - Stratified Grouped) & 0.918 $\pm$ 0.054 & 0.850 $\pm$ 0.003 \\
\hline
\hline
RF (Euclidean Agglomeration) & 1.0 $\pm$ 0.0 & 0.803 $\pm$  0.026 \\
\hline
RF (Cosine Agglomeration) & 1.0 $\pm$ 0.0 & 0.899 $\pm$ 0.002 \\
\hline
RF (Exemplar Agglomeration) & 1.0 $\pm$ 0.0 & 0.732 $\pm$ 0.007 \\
\hline
\hline
\textbf{SparseNN} & \textbf{1.0 $\pm$ 0.0}  &  \textbf{0.908 $\pm$ 0.007} \\
\end{tabular}
\caption{Summary of wall classification performance with different models and feature elimination criteria.}  
\label{tab:wall_classifcation_overall_performance_summary}
\end{table}

\subsubsection{Feature Robustness and Locations}
\label{sec:wall_classification_feature_robustness}
Similar to the stud detection task, one of the deeper SparseNN models (i.e with higher abstraction potential), specifically the (8,8,8,n) SparseNN configuration,was used to train 12 separate models with the same train-test split. This repetition aimed to assess the robustness of the trimmed feature sets selected during training.  

Nearly identical feature sets are obtained across models, with only four iterations showing slight deviations. The feature at 0.934 ns appears to be spurious or unimportant; its removal results in a minor performance boost, as shown in the bottom panel. In contrast, the feature at 1.685 ns is evidently critical; its exclusion, even when all other features are retained, leads to a 10\% drop in accuracy.


\begin{table}[H]
\centering
\small
\begin{tabular}{c|cccccccccccc}
\hline
\textbf{Features (ns)} & \textbf{1} & \textbf{2} & \textbf{3} & \textbf{4} & \textbf{5} & \textbf{6} & \textbf{7} & \textbf{8} & \textbf{9} & \textbf{10} & \textbf{11} & \textbf{12} \\
\hline
0.696 & \textbullet & \textbullet & \textbullet & \textbullet & \textbullet & \textbullet & \textbullet & \textbullet & \textbullet & \textbullet & \textbullet & \textbullet \\
0.715 & \textbullet & \textbullet & \textbullet & \textbullet & \textbullet & \textbullet & \textbullet & \textbullet & \textbullet & \textbullet & \textbullet & \textbullet \\
0.934 & \texttimes & \textbullet & \textbullet & \texttimes & \textbullet & \textbullet & \textbullet & \textbullet & \textbullet & \textbullet & \texttimes & \textbullet \\
1.173 & \textbullet & \textbullet & \textbullet & \textbullet & \textbullet & \textbullet & \textbullet & \textbullet & \textbullet & \textbullet & \textbullet & \textbullet \\
1.374 & \textbullet & \textbullet & \textbullet & \textbullet & \textbullet & \textbullet & \textbullet & \textbullet & \textbullet & \textbullet & \textbullet & \textbullet \\
1.685 & \textbullet & \textbullet & \textbullet & \textbullet & \texttimes & \textbullet & \textbullet & \textbullet & \textbullet & \textbullet & \textbullet & \textbullet \\
1.704 & \texttimes & \texttimes & \texttimes & \textbullet & \texttimes & \texttimes & \texttimes & \texttimes & \texttimes & \texttimes & \texttimes & \texttimes \\
\hline
\textbf{Accuracy} & 0.90 & 0.89 & 0.89 & 0.91 & 0.78 & 0.89 & 0.89 & 0.89 & 0.88 & 0.89 & 0.90 & 0.89 \\
\hline
\end{tabular}
\caption{Feature selection patterns across 12 SparseNN models. A filled dot (\textbullet) indicates that a feature (given by time in ns) was selected by the model, while a thin cross (\texttimes) indicates it was not. The final row shows the test accuracy of each model. Nearly identical feature sets are obtained across models, with only four iterations showing slight deviations. The feature at 0.934 ns appears to be spurious or unimportant; its removal results in a minor performance boost, as shown in the bottom panel. In contrast, the feature at 1.685 ns is evidently critical; its exclusion, even when all other features are retained, leads to a 10\% drop in accuracy.}
\label{tab:SparseNN_feature_robustness}
\end{table}


\begin{figure}[H]
    \centering
    \includegraphics[width = 1\linewidth]{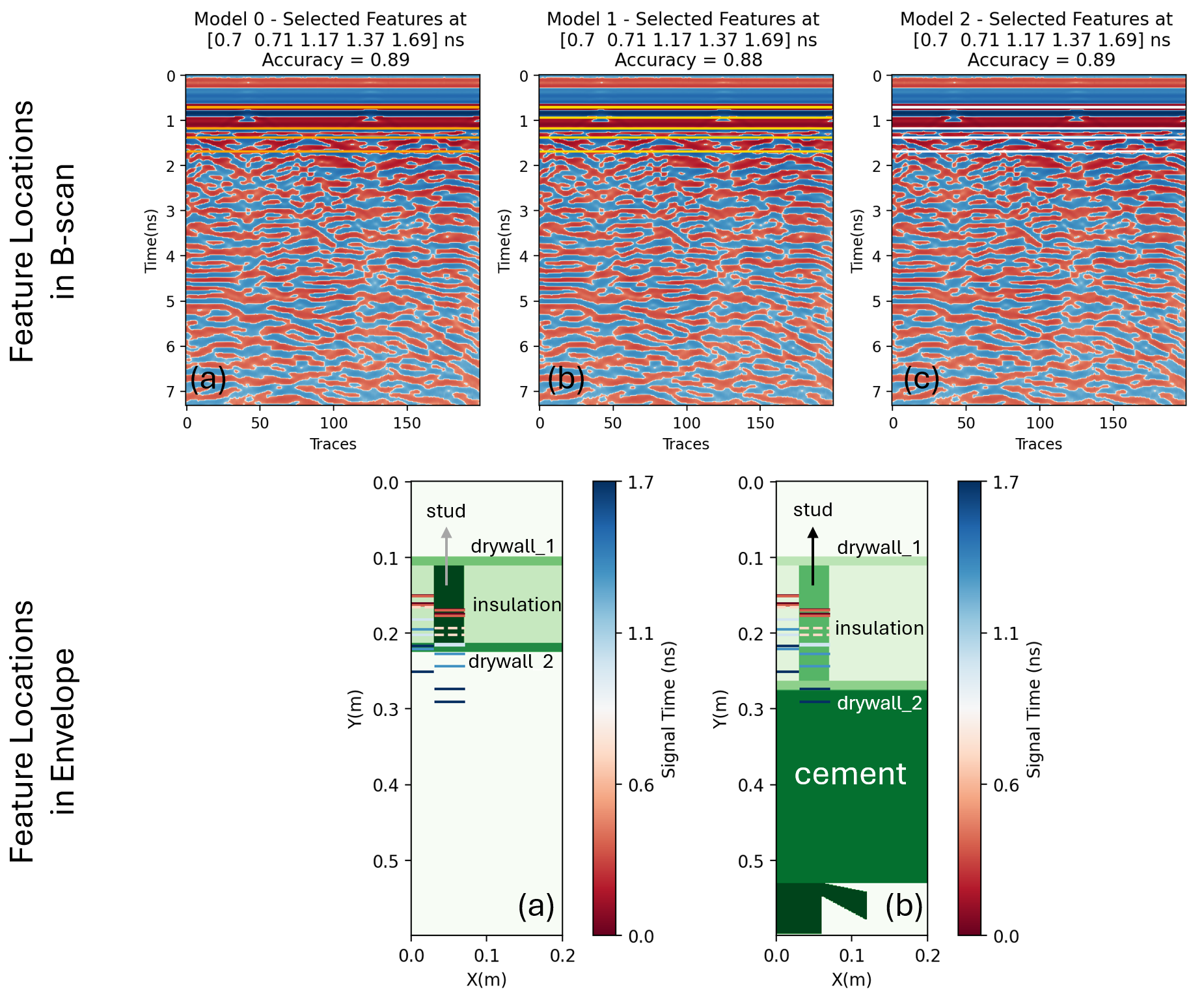}
 \caption{Feature stability and spatial interpretation of SparseNN wall-classification models. 
The top row \textbf{(a--c)} shows B-scans for three independently trained (8,8,8,n) SparseNN models, with selected feature locations marked as horizontal lines. 
Models~0 and~2 converge to identical feature sets at 0.70, 0.71, 1.17, 1.37, and 1.69~ns, achieving accuracies of~0.89, whereas Model~1 includes an additional feature at~0.93~ns, which slightly reduces accuracy to~0.88. 
The lower panels \textbf{(a--b)} map these temporal features to their possible spatial positions within the wall assemblies by converting signal times to depth using the minimum and maximum relative permittivities of each material layer. 
Each feature, therefore, appears as a pair of horizontal lines representing these bounds. 
The Red--Blue colormap denotes signal time, and the Y-axis indicates physical depth. 
The near-white dotted pair at~0.93~ns corresponds to the spurious feature observed in Model~1, which does not align with a material interface and degrades performance. 
Most other selected features fall within regions expected to vary between interior and exterior walls, underscoring the physical relevance and consistency of the learned representations.} 
    \label{fig:wall_classification_feature_locations}
\end{figure}

Figure~\ref{fig:wall_classification_feature_locations} shows the temporal and spatial locations of the features identified by repeated training of the SparseNN wall-classification models. 
The top panels display the selected feature times within the B-scan, while the lower panels map these temporal features to their corresponding spatial coordinates within the wall assemblies. 
The depth conversion is based on wave-propagation calculations using the minimum and maximum relative permittivities of each material layer (details in Appendix~\ref{sec:wave_propagation_calculation}). 
Each feature, therefore, appears as a pair of horizontal lines representing these permittivity bounds. 
The Red--Blue colormap denotes signal time, and the vertical axis ($Y~[\text{m}]$) indicates physical depth. 
Most features lie within regions expected to differ between interior and exterior walls, whereas the dotted near-white pair at~0.93~ns corresponds to a spurious feature that does not align with a material interface and leads to a slight deterioration in model performance.

\subsubsection{SHAP Analysis}

To further evaluate the interpretability of the SparseNN models, SHAP (SHapley Additive exPlanations) analysis was applied (see Section~\ref{sec:methods} for details). SHAP assigns each input feature a value indicating how strongly it contributes to increasing or decreasing the probability of the \emph{Exterior} class. In our binary setting of wall classification, positive SHAP values increase $P(\text{Exterior})$, while negative values decrease it, correspondingly increasing $P(\text{Interior})$.

\begin{figure}[H]
    \hspace{-1.3cm}
    \includegraphics[width = 1.2\linewidth]{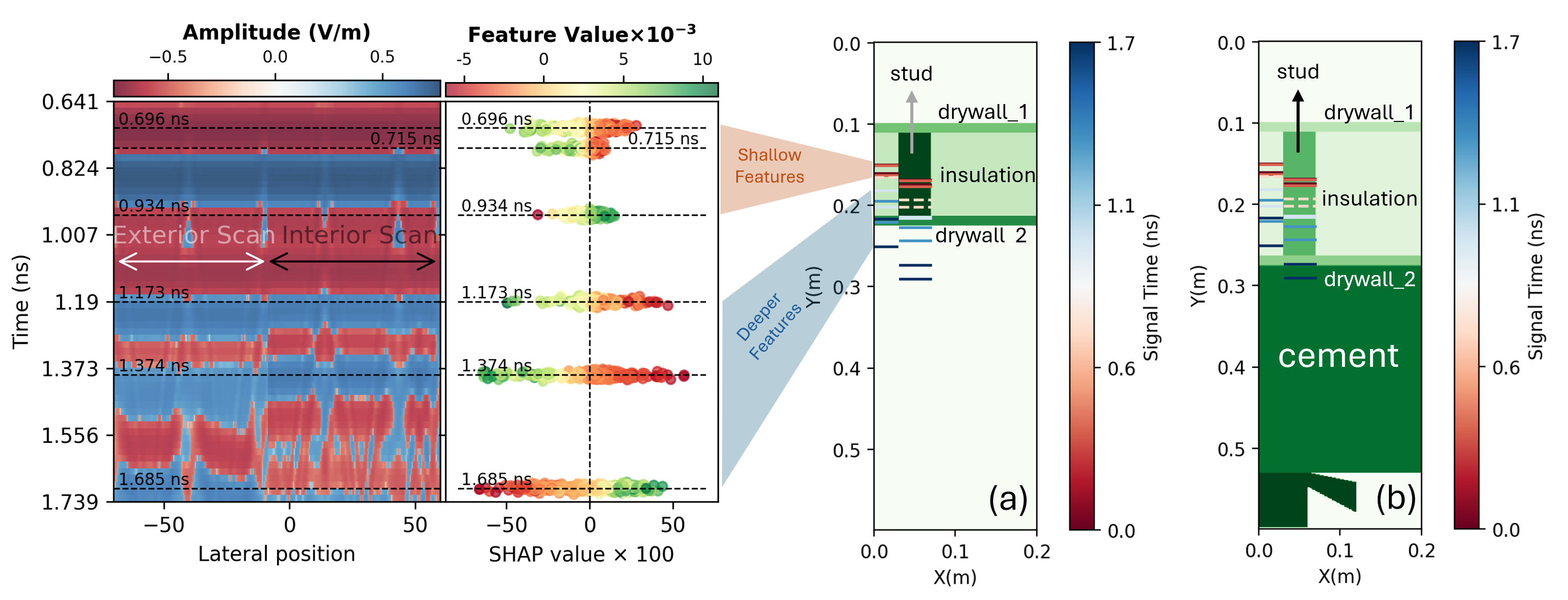}
    \caption{SHAP analysis of SparseNN-selected features. On the left, B-scan amplitudes are shown with dashed lines marking selected feature times, alongside the corresponding SHAP value distributions colored by signed feature values. Positive and negative reflections consistently bias predictions in opposite directions at a given time signal, with deeper features (1.173–1.685 ns) exhibiting the widest SHAP spreads. On the right, the same feature times are mapped onto the interior (a) and exterior (b) wall assemblies. The deeper SHAP features align with interfaces where material composition differs between interior and exterior walls, confirming that the most influential attributions correspond to meaningful dielectric boundaries. Shallow features (0.696–0.934 ns) show weaker but structured polarity effects, likely arising from systematic differences in antenna–wall coupling rather than discrete reflections.}
    \label{fig:shap_scatter_overlay}
\end{figure}

The SHAP plots in Figure~\ref{fig:shap_scatter_overlay} color each feature by its signed amplitude, capturing both sign (polarity) and strength. This reveals a clear pattern: at each time signal, positive-valued samples systematically bias predictions toward one class, while negative-valued samples bias them toward the other. In deeper regions of the scan (1.173--1.685~ns), these sign-dependent contributions are strongest and exhibit the widest spread of SHAP values. This is consistent with wave propagation computations (Figures~\ref{fig:wall_classification_feature_locations} and \ref{fig:shap_scatter_overlay}), which show that these features occur at material interfaces that differ between interior and exterior walls, such as insulation boundaries and cement backings. The alignment between SHAP contributions and structural interfaces confirms that the network is leveraging physically meaningful cues in its decision process.

At earlier times (0.696--0.934~ns), the SHAP distributions are narrower and weaker in magnitude, but they remain structured: positive and negative samples still bias predictions in consistent directions. These shallow bands do not correspond to major dielectric boundaries; instead, they likely capture system-level responses and differences in how the antenna couples into the wall stacks. For instance, interior walls with 2$\times$4 studs and exterior walls with 2$\times$6 studs could lead to variable higher-order reflections within the envelope, creating subtle but reproducible polarity biases even before the first distinct interface is reached.

In summary, SparseNNs discriminate wall type primarily by exploiting the sequence of signed reflection values associated with material interfaces. The strongest SHAP attributions align with deeper features where wall composition varies between interior and exterior assemblies, reinforcing the physical interpretability of the model. Shallow features contribute weaker but still systematic polarity effects linked to coupling differences. Overall, the network’s behavior mirrors radar physics: the sign change of reflection at dielectric contrasts drives the prediction direction, while the structural context determines where these cues are most influential.

\section{Conclusion}

This study addressed the challenge of interpreting ground-penetrating radar (GPR) data from building envelopes, where low dielectric contrast, compact multilayer assemblies, and subwavelength layer thicknesses lead to overlapping reflections and ambiguous signals. Within this setting, we analyzed GPR data through two linked classification tasks: stud detection, capturing lateral variations within the wall, and wall-type classification, capturing depth-dependent differences between interior and exterior assemblies.

Across both tasks, sparsity played a central role, enabling strong predictive performance while constraining the models to rely on a small subset of time-sample features. Random Forest baselines, together with variants incorporating recursive feature elimination, permutation-based importance measures, and feature agglomeration, provided competitive reference performance and a way to probe feature redundancy in the traces. In parallel, Sparse Neural Networks (SparseNNs) achieved comparable or higher accuracies while retaining only a small number of features, and repeated training runs converged to stable feature sets. Wave-propagation-based depth mapping and subsequent SHAP analyses showed that these retained features align with plausible material interfaces and regions where wall composition is expected to differ, indicating that the models are exploiting physically meaningful structure rather than arbitrary artifacts.

Taken together, these results show that sparsity-enforcing models, combined with restrained data requirements and post hoc physical interpretation, offer a viable path for data-driven GPR analysis of wall assemblies. The framework established here provides a baseline for extending GPR-based diagnostics beyond structural characterization of intact envelopes toward more complex tasks, including defect localization, tracking of degradation over time, and adaptation to a broader range of wall configurations and materials

\clearpage
\begin{center}
    \Large\bfseries Appendix
\end{center}
\vspace{1em}

\setcounter{section}{0}
\setcounter{subsection}{0}
\setcounter{figure}{0}
\setcounter{table}{0}

\renewcommand\thesection{\arabic{section}}
\renewcommand\thesubsection{\thesection.\arabic{subsection}}
\renewcommand\thefigure{\arabic{figure}}
\renewcommand\thetable{\arabic{table}}

\section{Exterior Wall Layout}

\begin{figure}[H]
    \centering
    \includegraphics[width = 1\linewidth]{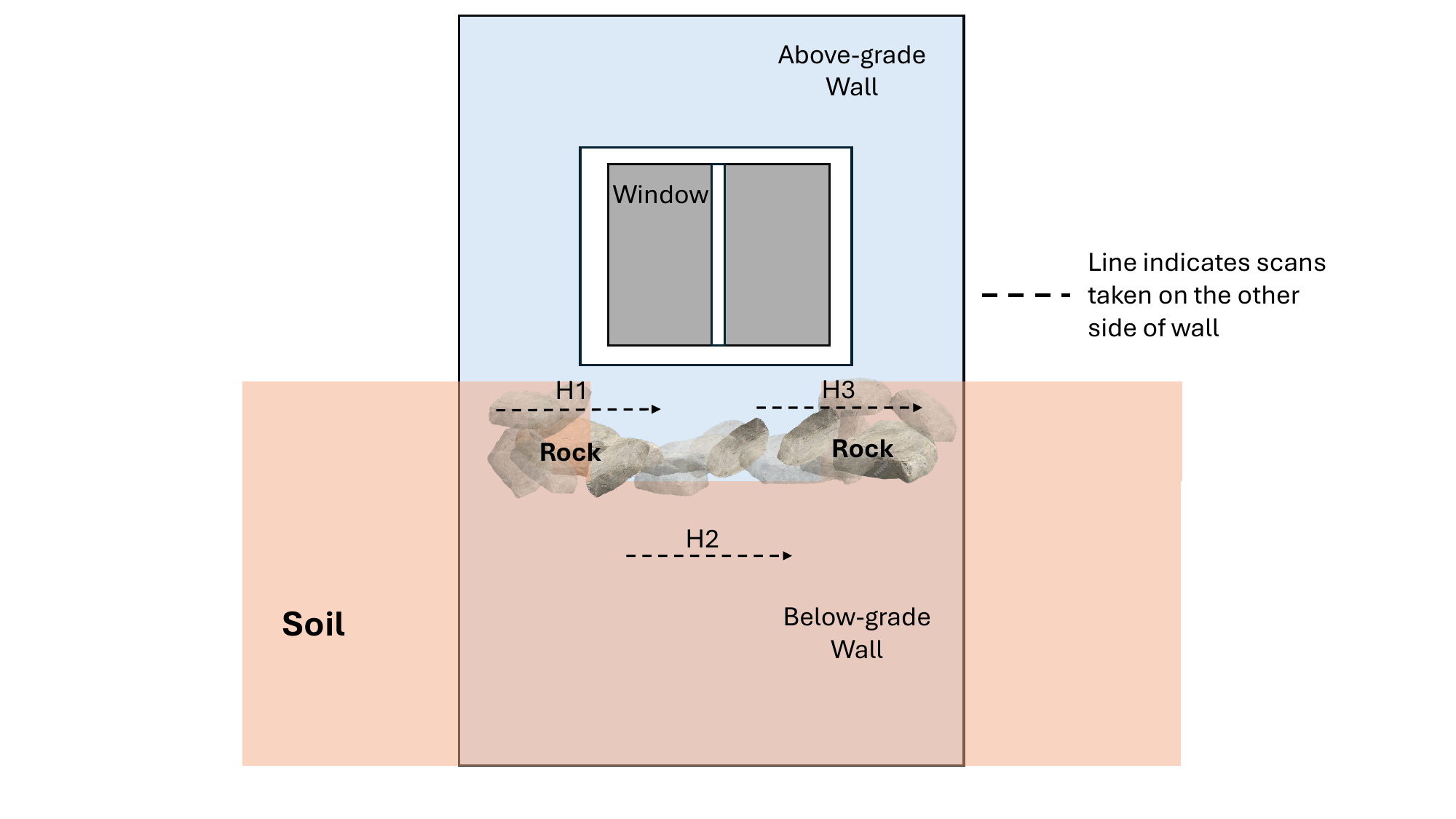}
    \caption{ Schematic of exterior wall with window wells that engulf portions of the exterior wall with rock and soil. As shown in the figure, exterior wall scan segments can have varying backgrounds. For instance, in this wall, \texttt{H2}'s background is completely soil, whereas \texttt{H1} and \texttt{H3}'s background consists of a varying rock/soil mixture gradually transitioning to air (space) in the middle of the wall.}
    \label{fig:window_well_schematic}
\end{figure}

\section{Stud Detection}

\subsection{SVD Cut-off Tuning}

\begin{figure}[H]
    \centering
    \includegraphics[width = 0.8\linewidth]{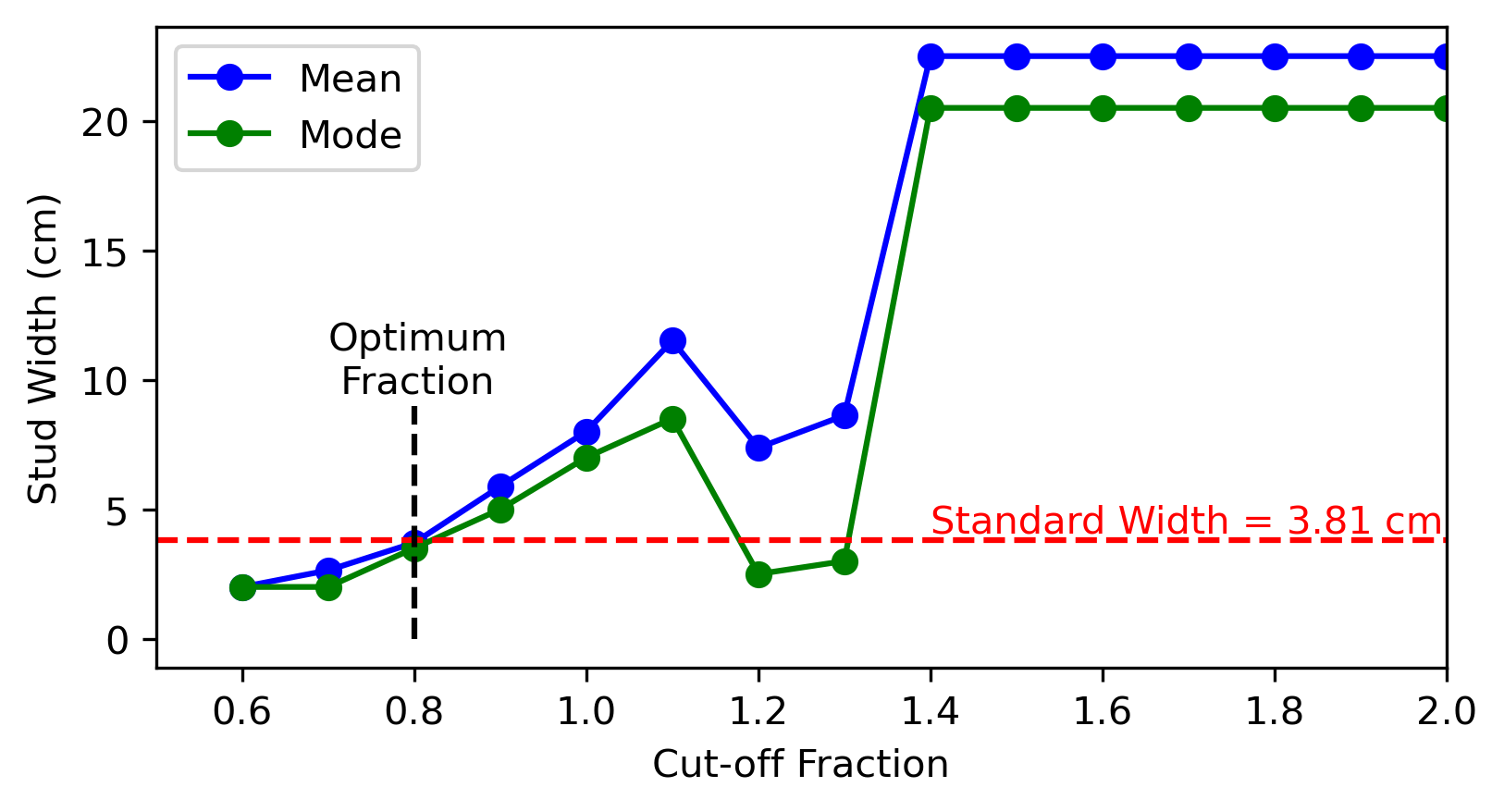}
    \caption{Tuning of the threshold fraction that defines stud. The red dotted line is the standard stud width of 1.5 inches. The cut-off fraction was chosen so that the mode of the stud width distribution coincided with 1.5 inches as closely as possible. This value of the cut-off fraction also yields a mean stud width close to 1.5 inches.}
    \label{fig:stud_detection_svd_cut_off}
\end {figure}

\begin{figure}[H]
    \centering
    \includegraphics{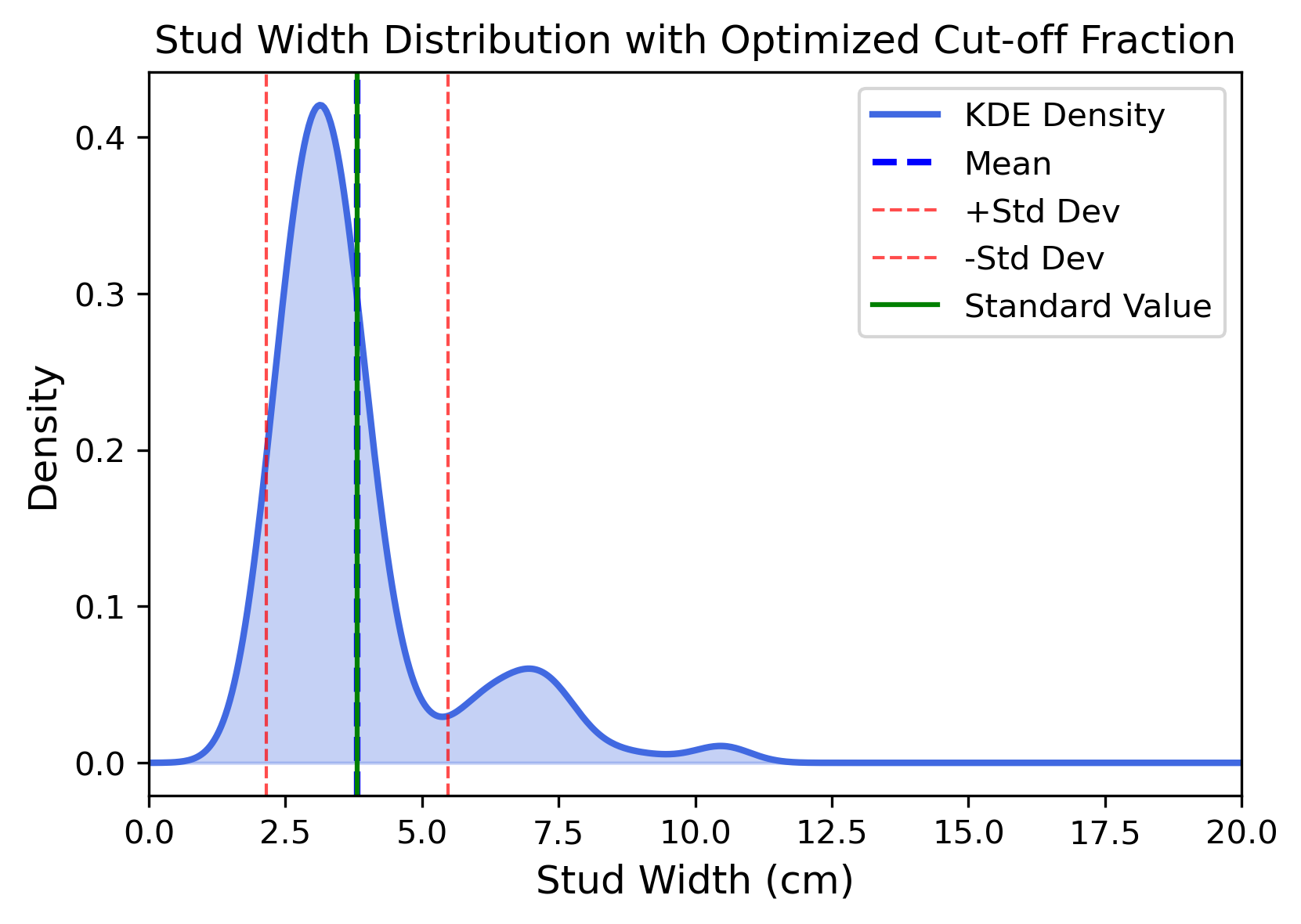}
    \caption{Stud width distribution with optimized threshold fraction. }
    \label{fig:stud_width-distribution}
\end{figure}

\subsection{Feature Minimization}
\label{sec:appendix_stud_detection_feature_minimization}

The ML models treat each time sample as an individual feature. Thus, many spurious and irrelevant features are present in the input data, arising from noise and nearby structures, surfaces, and inhomogeneities unrelated to studs. Moreover, there is a strong correlation among the building features. Therefore, several feature reduction techniques are employed and compared to generalize and improve the model's performance. 

\subsubsection{Similar-based Feature Selection - Feature Agglomeration}
\label{sec:appendix_stud_detection_feature_agglomeration}

Feature agglomeration is a dimensionality reduction technique that groups similar features using hierarchical clustering, with a specified distance metric (e.g., Euclidean or cosine) and linkage criterion (e.g., average or complete). In this study, features represent time-sampled amplitudes in A-scans, which are known to exhibit strong local correlations due to the temporal structure of reflected radar signals. By leveraging these correlations, feature agglomeration reduces redundancy by merging similar features into representative clusters, effectively compressing the input dimensionality while possibly preserving key signal variations. This not only improves model efficiency and training speed but also helps mitigate overfitting by suppressing high-frequency noise and irrelevant fluctuations.

In this study, three feature agglomeration strategies were implemented to investigate how different similarity measures and clustering choices affect dimensionality reduction and predictive performance. The first two approaches—Euclidean and cosine distance-based agglomeration—utilize hierarchical clustering followed by mean pooling, where each cluster of correlated features is replaced by the group's average signal. Euclidean distance captures absolute differences between time samples, making it sensitive to amplitude, while cosine distance reflects angular similarity and emphasizes waveform shape. 

The third strategy—exemplar-based agglomeration—eschews pooling entirely. Instead, it selects a single feature from each cluster (typically the one closest to the cluster's centroid) as a representative. While this approach leads to lower classification accuracy than pooled agglomeration, it retains the physical traceability of input features and avoids interpolation artifacts introduced by averaging. More importantly, it provides a consistent comparative framework across the paper: because exemplar agglomeration retains a strictly filtered version of the original input, it allows for a direct and interpretable comparison between model performance and the number of features used across different dimensionality reduction and modeling strategies. As such, exemplar-based agglomeration serves as a reference baseline for evaluating the performance-to-feature ratio of more complex pipelines such as SparseNN and Recursive Feature Elimination.

\begin{figure}[H]
    \centering
\includegraphics{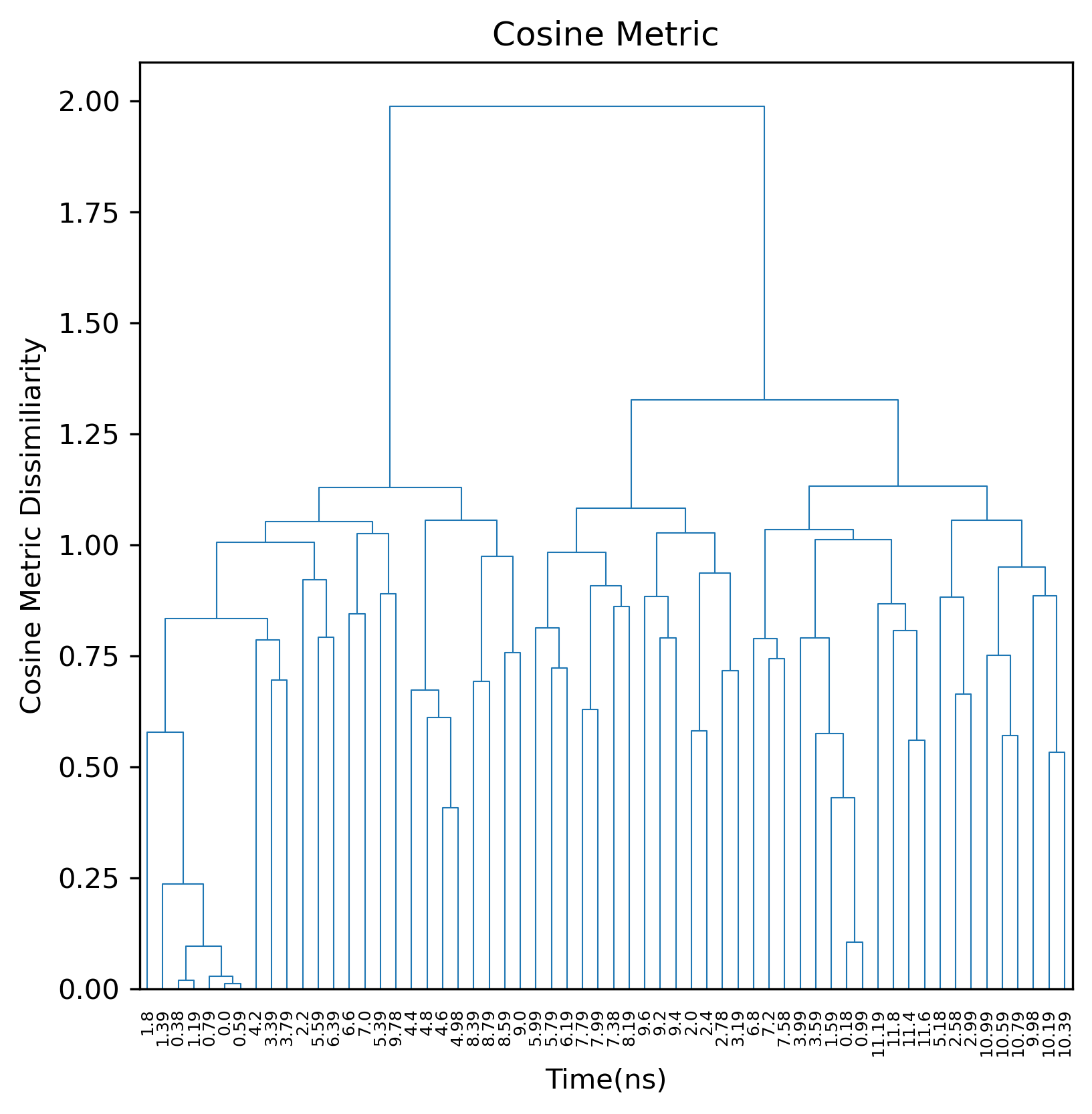}
    \caption{A visual representation of the feature agglomeration process. Nearby signals in time are correlated in the A-scans. Feature agglomeration uses this attribute to cluster the input feature points and reduce their dimension from 655 to 2. In this figure, the number of features shown is sub-sampled by a factor of  10 for ease of depiction.}
    \label{fig:agglomerated_features}
\end{figure}

For each agglomeration strategy, models were trained with 1 to 50 feature clusters to explore a wide range of dimensionality reduction. To account for stochastic variability in both clustering and model training, each configuration was repeated 10 times, and the mean accuracy and standard deviation (error bars) are reported. These results are summarized in Figure~\ref{fig:agglomerated_features}, which illustrates the trade-off between feature count and model accuracy for each method.

Among the three methods, Euclidean agglomeration performs best, achieving an accuracy comparable to that obtained using all features ($\approx 0.965$) while requiring only 10--15 pooled features and exhibiting minimal variance across repeated runs. In contrast, exemplar-based agglomeration, which avoids pooling and retains individual features from each cluster, shows a non-monotonic performance trend: it reaches competitive accuracy ($\approx 0.93$) with as few as 3--4 features, then drops in performance as more clusters are introduced, before recovering again around 30 clusters. This dip may be attributed to the inclusion of partially redundant or noisy features that dilute the signal quality without the smoothing effect of pooling. At higher cluster counts, the improved spatial coverage appears to restore discriminative power. Despite its variability, the exemplar method provides a useful lower-bound reference for evaluating sparsity-performance tradeoffs, as it preserves physical traceability of input features and avoids signal averaging. Cosine-based agglomeration, though effective in wall classification tasks (as seen in Section~\ref{sec:wall_classification}), underperforms here---plateauing around $88\%$ accuracy regardless of cluster count---likely due to its insensitivity to amplitude variations that are critical for identifying stud reflections. The strong performance of Euclidean agglomeration and SparseNN demonstrates that high-accuracy stud detection is achievable using compact, low-dimensional representations, thereby reducing computational cost while preserving fidelity.

\begin{figure}[H]
    \centering
\includegraphics[width = 0.9\textwidth]{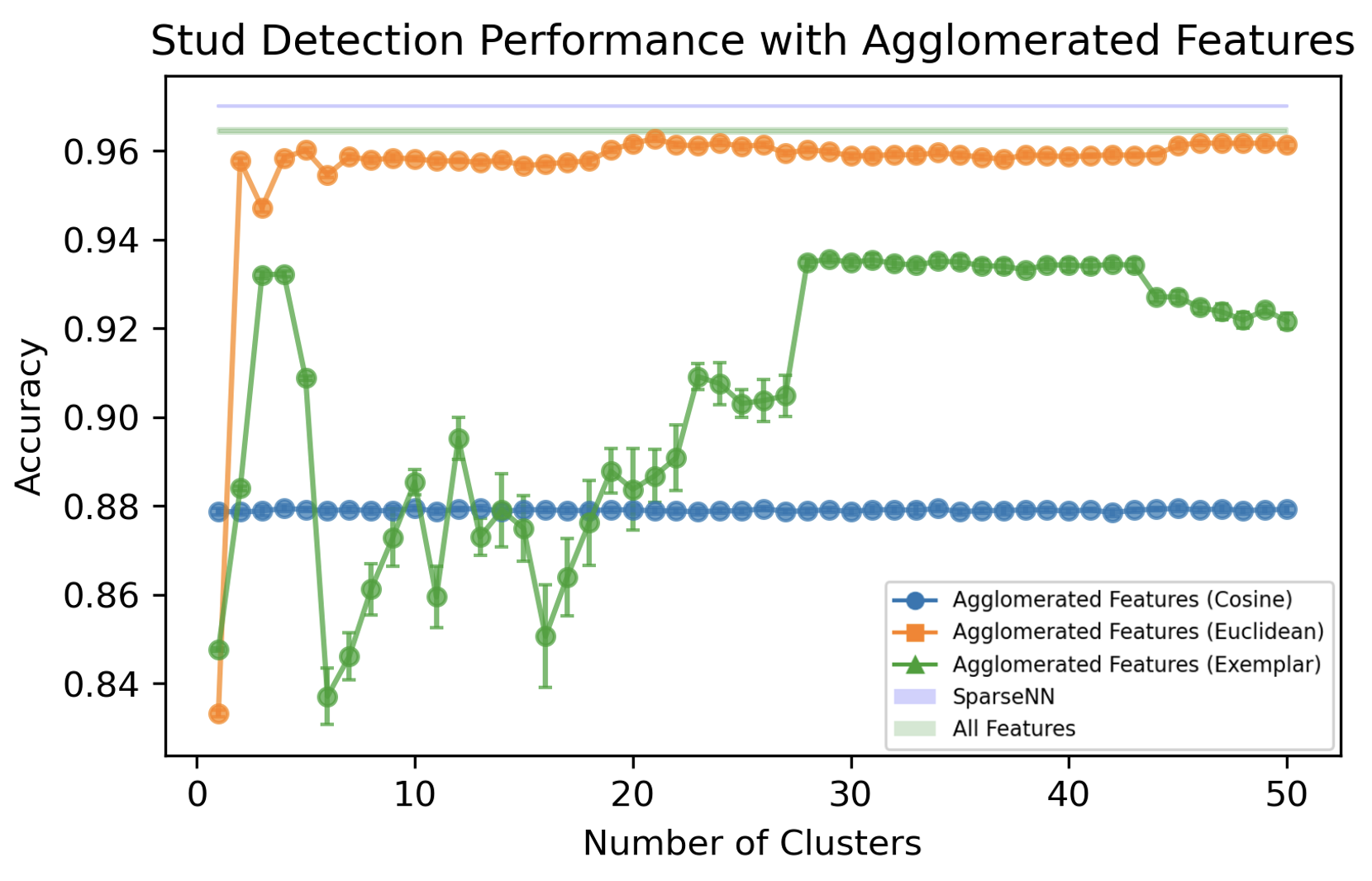}
    \caption{Stud prediction with agglomerated features, using Euclidean distance – only three feature clusters achieve performance that is on par with prediction performance using all 655 features in each trace. Optimized SparseNN (architecture: (8,); $\lambda_{reg} = 10^{-5}$) performance is significantly superior.}
\label{fig:stud_detection_agglomerated_features_performance}
\end{figure}

\subsection{Performance-based Feature Selection}
\label{sec:appnedix_stud_detection_performance_based_feature_selection}
\subsubsection{Permutation Feature Importance}
\label{sec:appendix_stud_detection_pfi}

While similarity-based feature agglomeration reduces dimensionality by clustering redundant or correlated inputs, performance-based feature selection takes a task-driven approach. Rather than relying on pairwise feature similarity, it directly assesses each feature's contribution to the model’s predictive performance. In this study, we initially use permutation feature importance (PFI), in which features are shuffled one at a time across the test set, disrupting their relationships with the target while keeping other features intact. For each feature, this process is repeated 10 times using different permutations, and the average drop in accuracy relative to the unshuffled baseline is recorded as the importance score. Features that cause negligible performance degradation when shuffled are deemed uninformative and may be safely excluded. This approach provides a complementary perspective to agglomeration by explicitly identifying the most discriminative regions in the signal, enabling interpretable sparsity.

\begin{figure}[H]
    \centering
\includegraphics{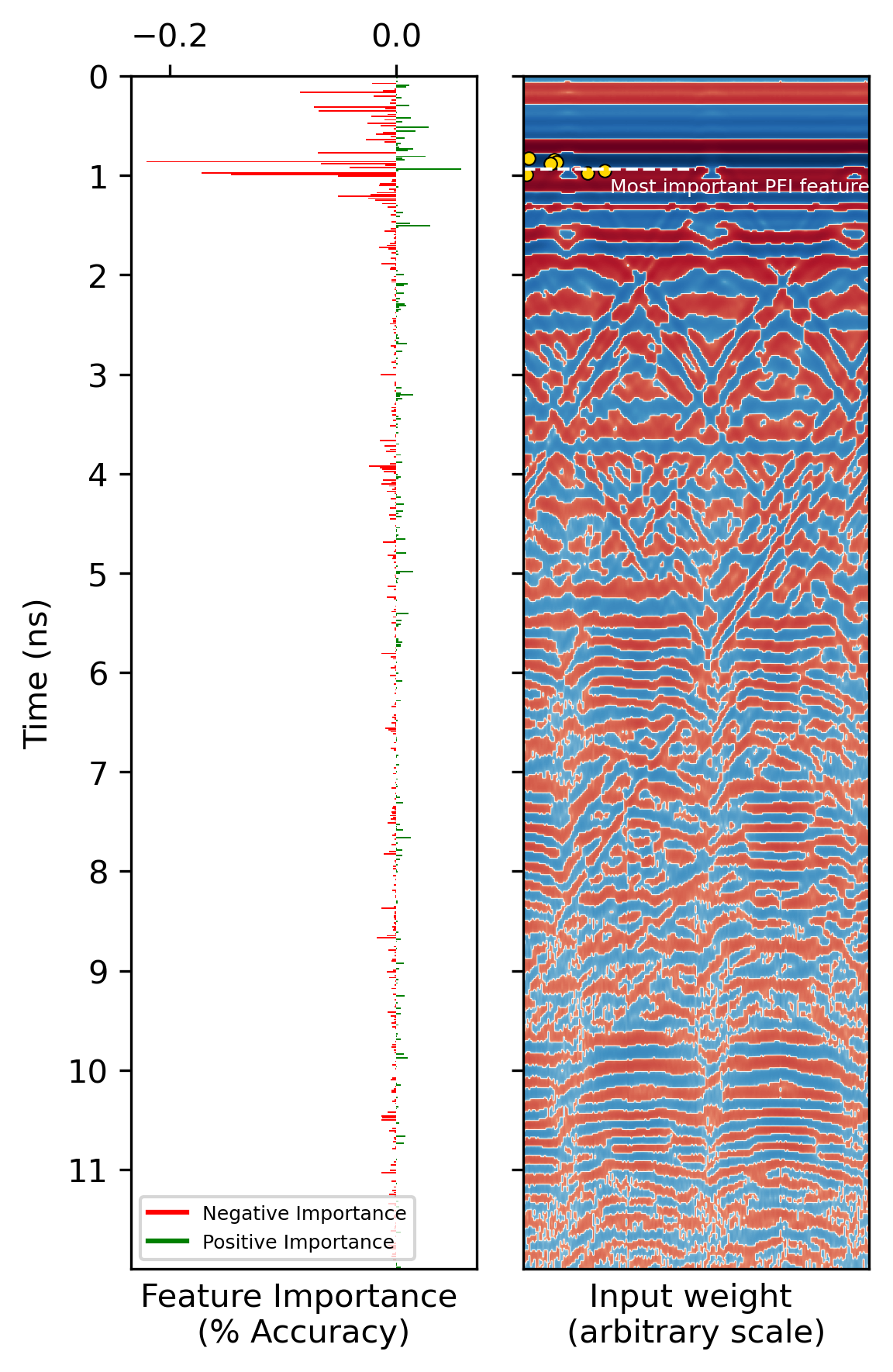}
    \caption{Feature importances for the stud detection task. Features marked in red have negative importance, indicating that randomly shuffling them results in performance gains. Features marked green are important, indicating that randomly shuffling them leads to performance deterioration. Thus, green features are likely correlated with the presence of studs. The location of the features also coincides with the pruned features with SparseNN (architecture: (8,n); $\lambda_{reg} = -5$)}
\label{fig:PFI_stud_detection}
\end{figure}


In  Figure~\ref{fig:PFI_stud_detection}, the PFI scores are shown in alignment with a sample B-scan from the dataset. The left panel shows the PFI scores, whereas the right panel shows the B-scan, with the most important features obtained by PFI and SparseNN, overlayed. Only a small band of features - primarily concentrated near 1~ns - exhibit relatively high importance, with even the strongest features contributing a modest but consistent drop in accuracy (up to $\approx 0.2\%$) when permuted. In contrast, a majority of features across the trace yield negligible or negative importance, particularly beyond 2~ns, when far-field clutter is likely to dominate the signal. The alignment between the most important PFI features and the SparseNN-selected indices suggests that sparse models not only achieve competitive accuracy but also learn to focus on physically meaningful signal regions. 

\subsubsection{Recursive Feature Elimination}
\label{sec:appendix_stud_detection_rfe}

While PFI provides a global ranking of individual feature importances, it evaluates each feature independently and does not account for feature redundancies or interactions. For instance, if two features are highly correlated, shuffling one may not lead to a significant drop in model performance, as the other can still carry the signal. This limitation can obscure the importance of redundant yet useful features.
To address this limitation, we applied Recursive Feature Elimination with Cross-Validation (RFECV) to identify a performance-optimized subset of features. RFECV iteratively removes the least important features, as determined by their mean decrease in Gini impurity from the Random Forest model, retraining the model after each removal. By re-evaluating feature importance at every step, the algorithm captures multivariate dependencies and progressively refines the feature set. The optimal subset is then selected based on the highest mean cross-validated accuracy, and the final model is evaluated on the held-out test set to assess generalization.

To ensure robust and representative evaluation, we adopt a stratified cross-validation strategy. Stratification preserves the proportion of classes (i.e., presence or absence of a stud) in each fold, which is essential for balanced training and evaluation. Two variants of this strategy are employed in our study: (Strategy 1) standard Stratified K-Fold, where samples are randomly split across folds while maintaining class balance, and (Strategy 2) Stratified Group K-Fold, which additionally enforces group-level consistency by ensuring that all samples from a given scan are assigned to the same fold. This prevents information leakage across scans from the same physical segment. Both strategies assess the stability and generalization of selected features under different evaluation constraints.




\paragraph{Strategy 1: Stratified Cross-validation}

This strategy divides the data samples into \( K \)-folds (\( K = 5 \) in this case), with each fold preserving the ratio of stud and non-stud labels. The dataset used for this procedure consists of traces from scans \texttt{G3} and \texttt{I1}. These traces are partitioned into five stratified folds. In each iteration, one fold serves as the validation set while the remaining four folds are used for training. During training, feature elimination is performed using a Random Forest model, where the five least important features, based on Gini importance, are removed at each iteration. The model is retrained after each elimination step, and validation accuracy is recorded to determine the optimal number of features. To account for variability due to model initialization and training randomness, the entire process is repeated ten times using the final selected feature set, and the mean and standard deviation of the resulting test accuracies are reported. Test accuracy is computed on traces from wall segments outside the training/validation set, that is, on scans other than \texttt{G3} and \texttt{I1}, to evaluate generalization to unseen envelope conditions. The test sets are used as-is, without any stratification.

\begin{table}[H]
\centering
\begin{tabular}{ c c c }
\hline
\textbf{Configuration} & \textbf{Train} & \textbf{Validation} \\
\hline
1-5 & Segments of \texttt{I1} and \texttt{G3}  & Left-out segments of \texttt{I1} and \texttt{G3} \\
\hline
\end{tabular}
\caption{Train and validation set for different folds in stratified RFECV.}  
\label{tab:stud_stratified_cv_split}
\end{table}

\begin{figure}[H]
    \centering
\includegraphics[width = 0.8\linewidth]{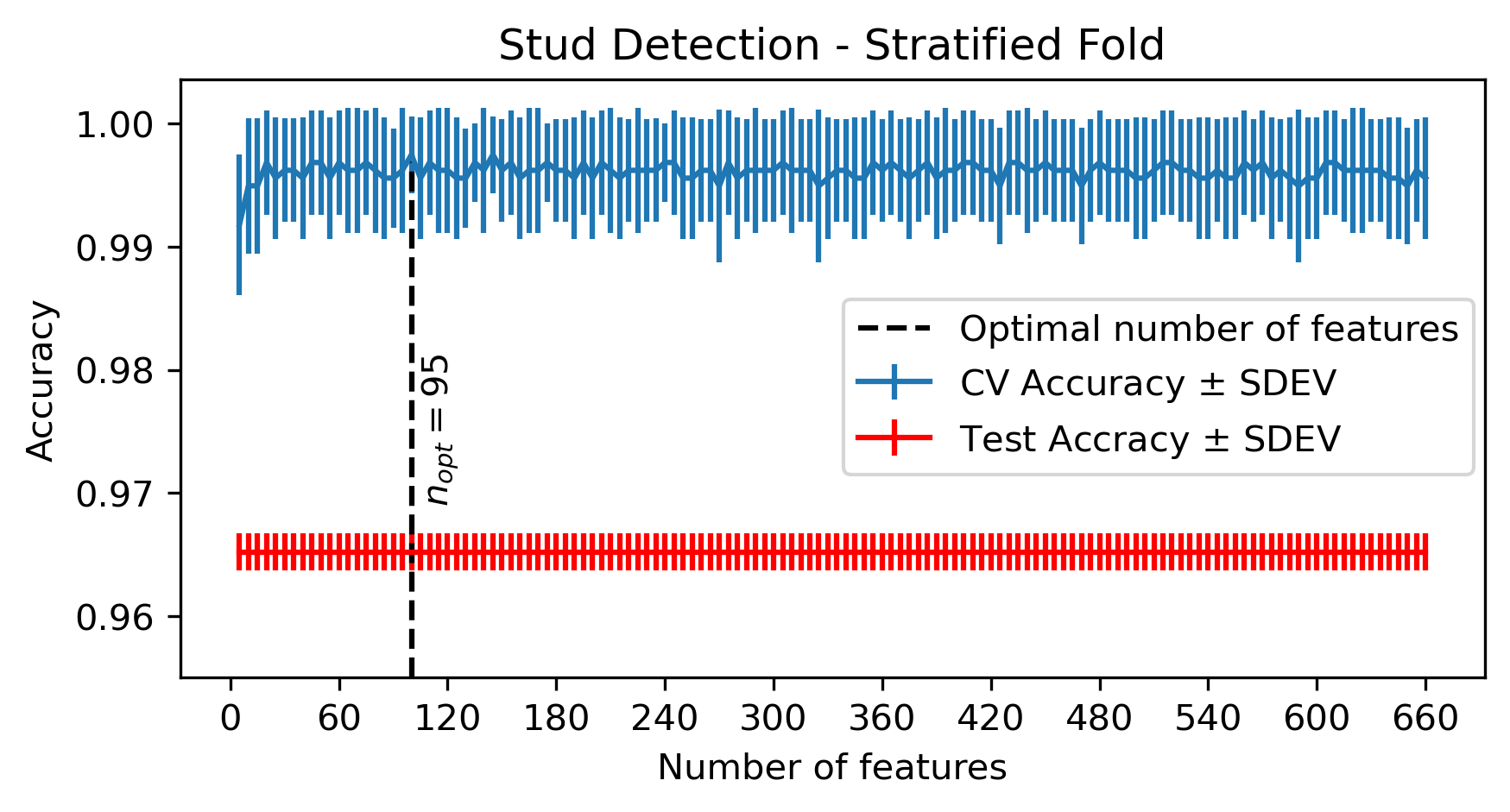}
    \caption{Recursive Feature Elimination with stratified cross-validation applied to stud detection using a minimal dataset (scans \texttt{I1} and \texttt{G3}). Validation accuracy exceeds 0.99 with as few as 27 features and peaks at 0.995 with 95 features. Test accuracy - evaluated on held-out scans not included in the training or validation folds—is 0.966. Training accuracy (not shown) reaches 1.0.}
\label{fig:RFI_stratified_stud_detection}
\end{figure}

Figure~\ref{fig:RFI_stratified_stud_detection} shows that cross-validation accuracy peaks at 0.995 with only 95 features, indicating that much of the input trace contributes little to performance. Test accuracy, evaluated using only this optimized feature subset on unseen wall segments, reaches approximately 0.966 and remains consistent across ten repeated runs (as indicated by the relatively small standard deviation). By computing test accuracy only at the selected feature count, we preserve the independence of the held-out test set. These results indicate the sparsity of meaningful information in the A-scan signal and motivate the development of models that operate effectively on compact feature representations.

\paragraph{Strategy 2: Stratified Group Cross-validation}

Stratified group cross-validation extends standard stratified cross-validation by ensuring that all samples from the same group—here, each segment scan—are assigned to the same fold. This prevents information leakage due to intra-scan correlations, as traces within a scan are spatially adjacent and may share low-level patterns unrelated to the target label. At the same time, the method aims to preserve the overall balance between stud and non-stud labels across folds. In this study, scans \texttt{I1}, \texttt{G2}, \texttt{B2}, and \texttt{D3} are treated as separate groups and used in the stratified group splitting process.

\begin{table}[H]
\centering
\begin{tabular}{ c c c }
\hline
\textbf{Configuration} & \textbf{Train} & \textbf{Validation} \\
\hline
1 & \texttt{G3},\texttt{B2} and \texttt{D3} & \texttt{I1} \\
\hline
2 & \texttt{I1}, \texttt{B2} and  \texttt{D3} & \texttt{G3} \\
\hline
3 & \texttt{I1} and \texttt{G3} & \texttt{B2} and \texttt{D3} \\
\hline
\end{tabular}
\caption{Train and validation set for different folds}  
\label{tab:stud_cv_stratfied_grouped_train_test_split}
\end{table}

\begin{figure}[H]
    \centering
    \includegraphics[width = 0.8\linewidth]{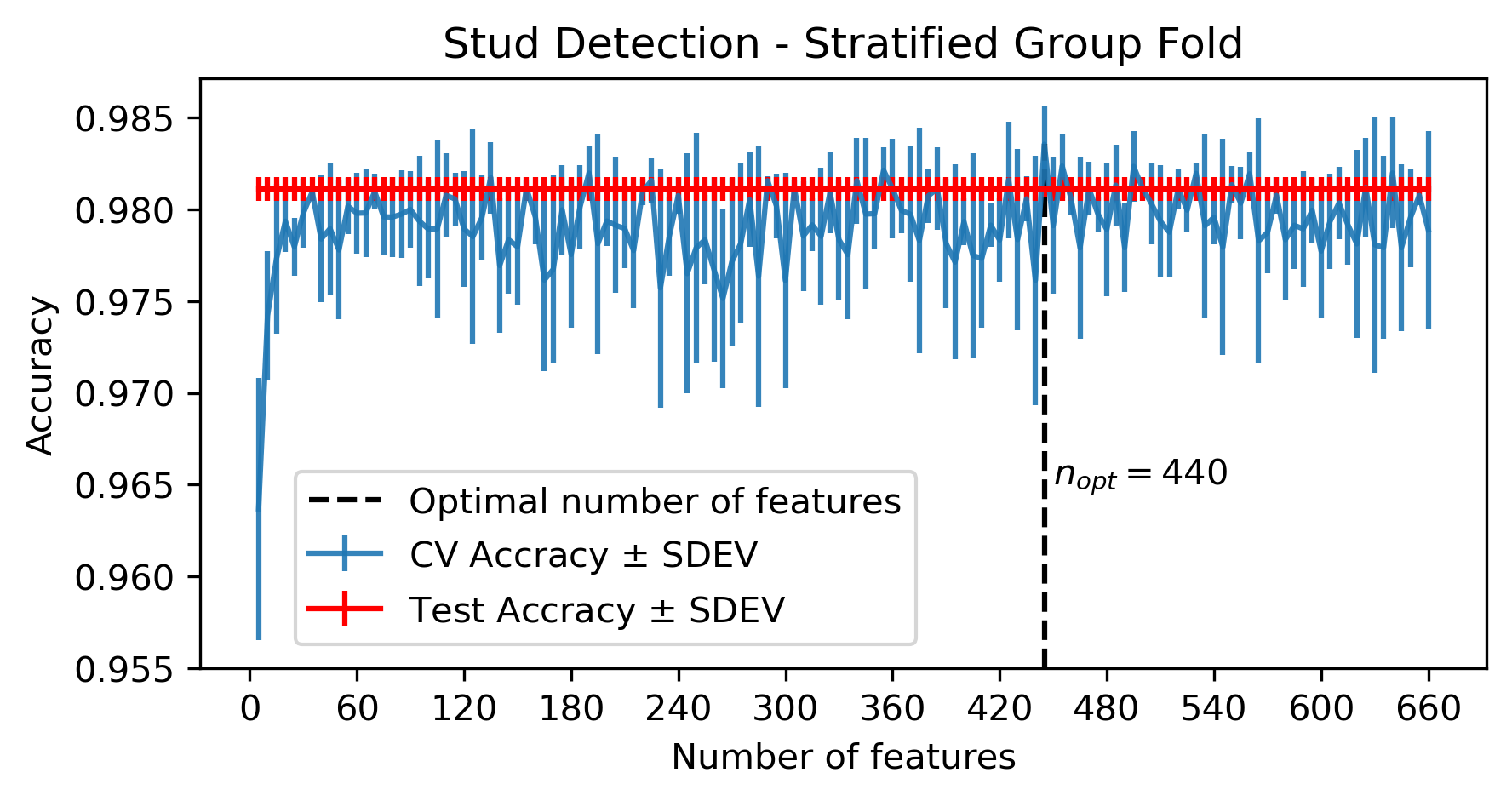}
    \caption{ Recursive Feature Elimination with stratified group cross-validation for stud detection using scans \texttt{I1}, \texttt{G3}, \texttt{B2}, and \texttt{D3}. Three train-validation configurations are used, as detailed in Table~\ref{tab:stud_cv_stratfied_grouped_train_test_split}. While peak validation accuracy (0.984) is reached with 440 features, near-optimal performance is already achieved with only 33 features. Test accuracy, evaluated on scans excluded from cross-validation, is 0.981. Training accuracy (not shown) reaches 1.000.}
    \label{fig:RFI_stratified_grouped_stud_detection}
\end{figure}

Figure~\ref{fig:RFI_stratified_grouped_stud_detection} shows the performance of RFECV using stratified group cross-validation. Compared to the previous stratified fold setup, cross-validation accuracy here is more variable but still peaks at approximately 0.984 with 440 features. This higher optimal feature count reflects the conservative nature of group-wise splitting, which introduces greater variation across folds due to scan-level separation. The corresponding test accuracy, evaluated on unseen scans and averaged across ten runs, is 0.981. This marks a notable improvement over the 0.966 achieved in the standard stratified fold setting. While the selected optimum feature set is relatively large, near-peak performance is already reached with as few as 33 features, indicating that most features contribute little beyond that point. These results reaffirm the sparsity of meaningful information in the signal and demonstrate that high accuracy is achievable with a compact, well-chosen subset of inputs.

\clearpage

\subsubsection{SparseNN Training}

\begin{figure}[!h]
    \centering  \includegraphics{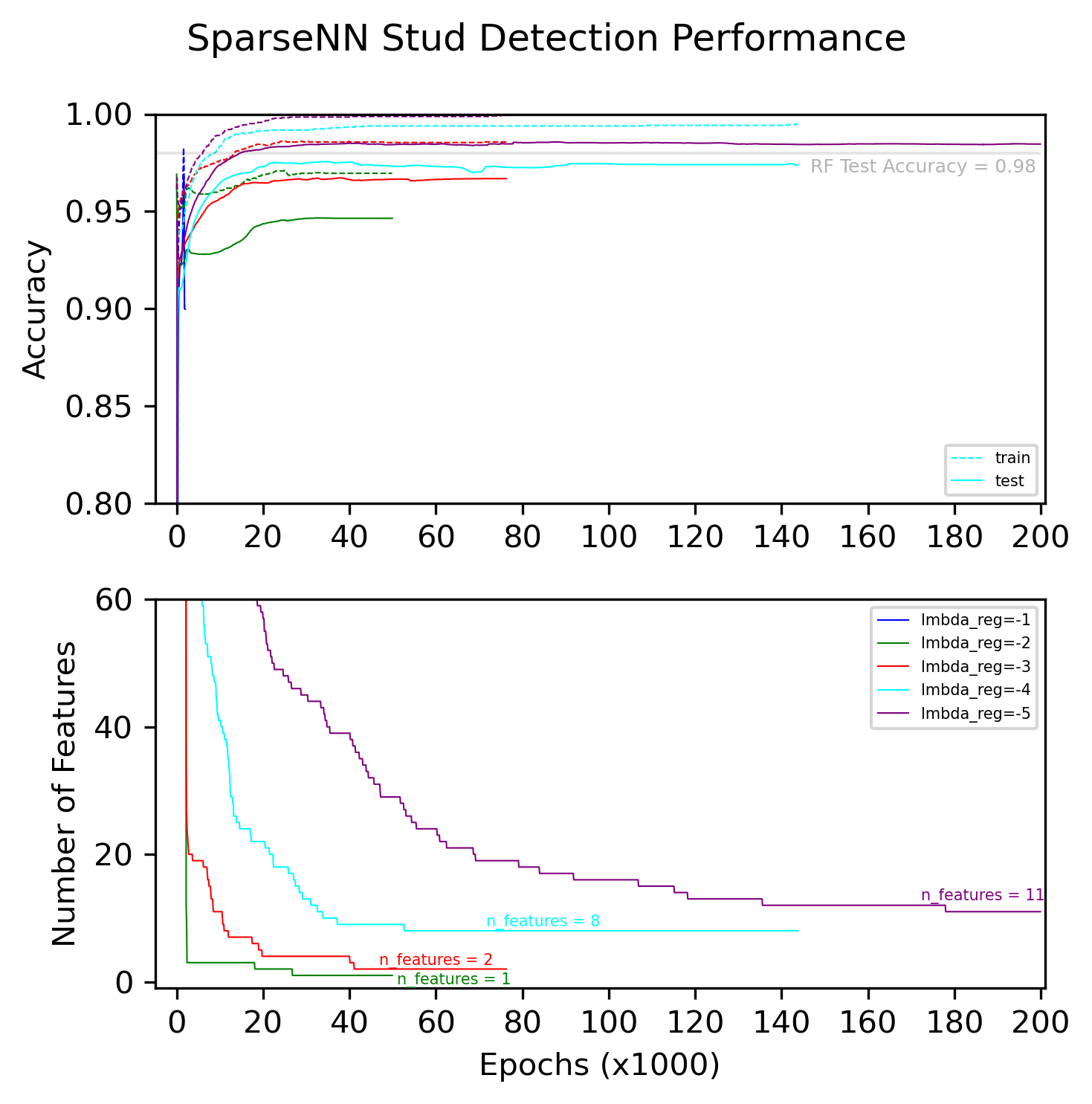}
    \caption{Performances of single-layer SparseNN with 8 hidden neurons for the stud detection task. Using only one input feature (green line), the model achieves accuracy reasonably close to that of a Random Forest trained on all features (approximately 0.93 vs. 0.98). The optimized SparseNN configuration (purple line) surpasses the RF benchmark, demonstrating the effectiveness of sparsity-driven feature selection.}
\label{fig:stud_SparseNN_performance}
\end{figure}

\clearpage

\subsubsection{Wave propagation–based calculations.}
\label{sec:wave_propagation_calculation}

\begin{figure}[H]
    \centering
    \includegraphics[width=1\linewidth]{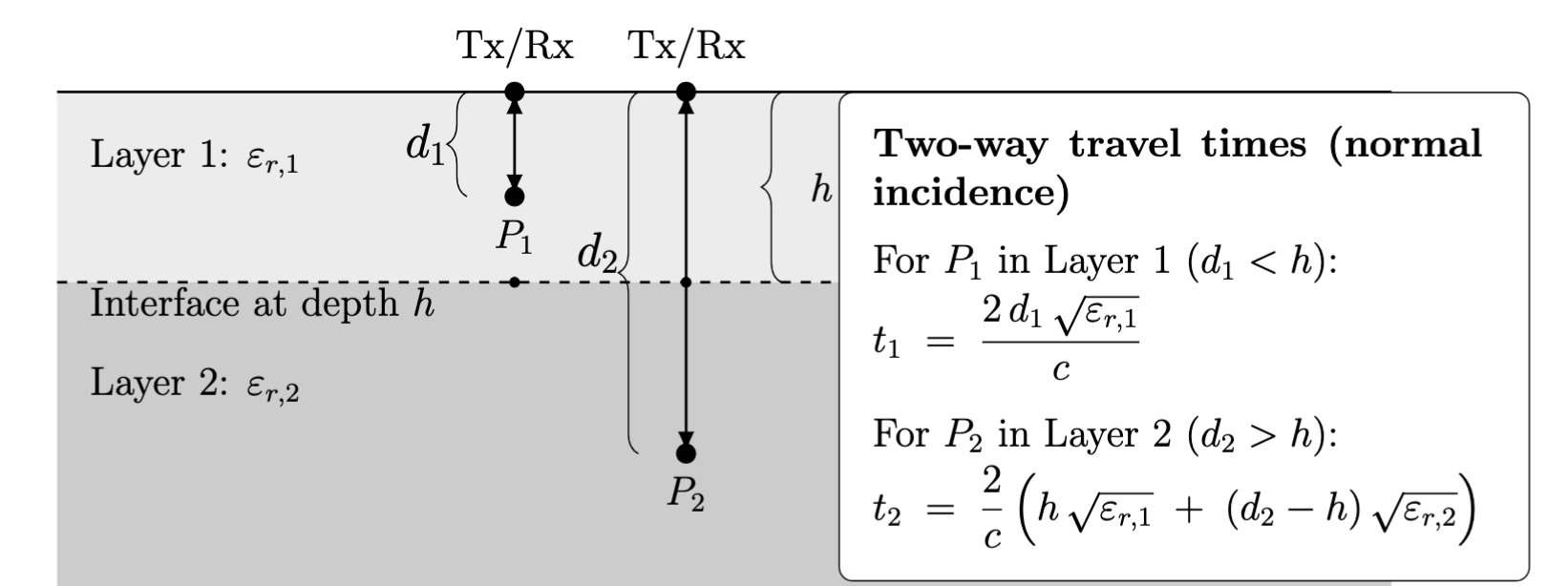}
    \caption{Wave propagation schematic for two-layer configuration with minimally separated Tx/Rx pairs (reflecting the field data acquisition setup) so each path is quasi-vertical. Reflector $P_1$ lies in Layer~1 ($d_1<h$); reflector $P_2$ lies in Layer~2 ($d_2>h$). Layer shading distinguishes media; braces for $d_1$ and $d_2$ sit just left of their columns, while $h$ is annotated close to the P2 column. Closed-form two-way times appear in a panel aligned with the top layer and shifted left to match the tighter layout.}
    \label{fig:wave_propagation_calculations}
\end{figure}

The Proceq GP8800 uses a near-offset (near-coincident) monostatic probe, in which the transmitter and receiver are housed together; the device’s compact form factor (8.9 cm cube; antenna-to-edge 4.5 cm) justifies the quasi-vertical path approximation, meaning that wave refraction and lateral travel components can be neglected. Under this assumption, the two-way travel times shown in Figure~\ref{fig:wave_propagation_calculations} can be expressed analytically for different reflector depths. For a target $P_1$ located within the first layer ($d_1 < h$), the pulse travels a round-trip distance of $2d_1$, giving $t_1 = 2d_1\sqrt{\varepsilon_{r,1}}/c$. For a deeper target $P_2$ lying below the interface ($d_2 > h$), the signal first traverses the upper layer and then continues through the lower medium, yielding
\[
t_2 = \frac{2}{c}\big(h\sqrt{\varepsilon_{r,1}} + (d_2 - h)\sqrt{\varepsilon_{r,2}}\big).
\]
In this work, these equations are used to map selected time-domain features to their corresponding spatial locations in the wall geometry under the quasi-vertical assumption, providing a direct link between signal behavior and physical structure.

\section{Wall Classification}

\subsection{Feature Minimization}
\label{sec:appendix_wall_classification_feature_minimization}
As mentioned earlier in Section~\ref{sec:methods} the ML models treat each time sample as an individual feature. Thus, many spurious and irrelevant features are present in the input data, arising from noise and nearby structures, surfaces, and inhomogeneities unrelated to studs. Moreover, there is a strong correlation among the building features. Therefore, several feature reduction techniques are employed and compared to generalize and improve the model's performance. 

\subsubsection{Similarity Based Feature Selection: Feature Agglomeration}

Unlike stud detection, agglomerated features (with the cosine metric lead to a considerable performance boost for wall classification. Unlike the case for stud detection, the features corresponding to wall type are more distributed throughout the A-scan. Thus, the feature distillation using agglomeration improves model performance appreciably. 

\begin{figure}[H]
    \centering
\includegraphics[width = 0.9\textwidth]{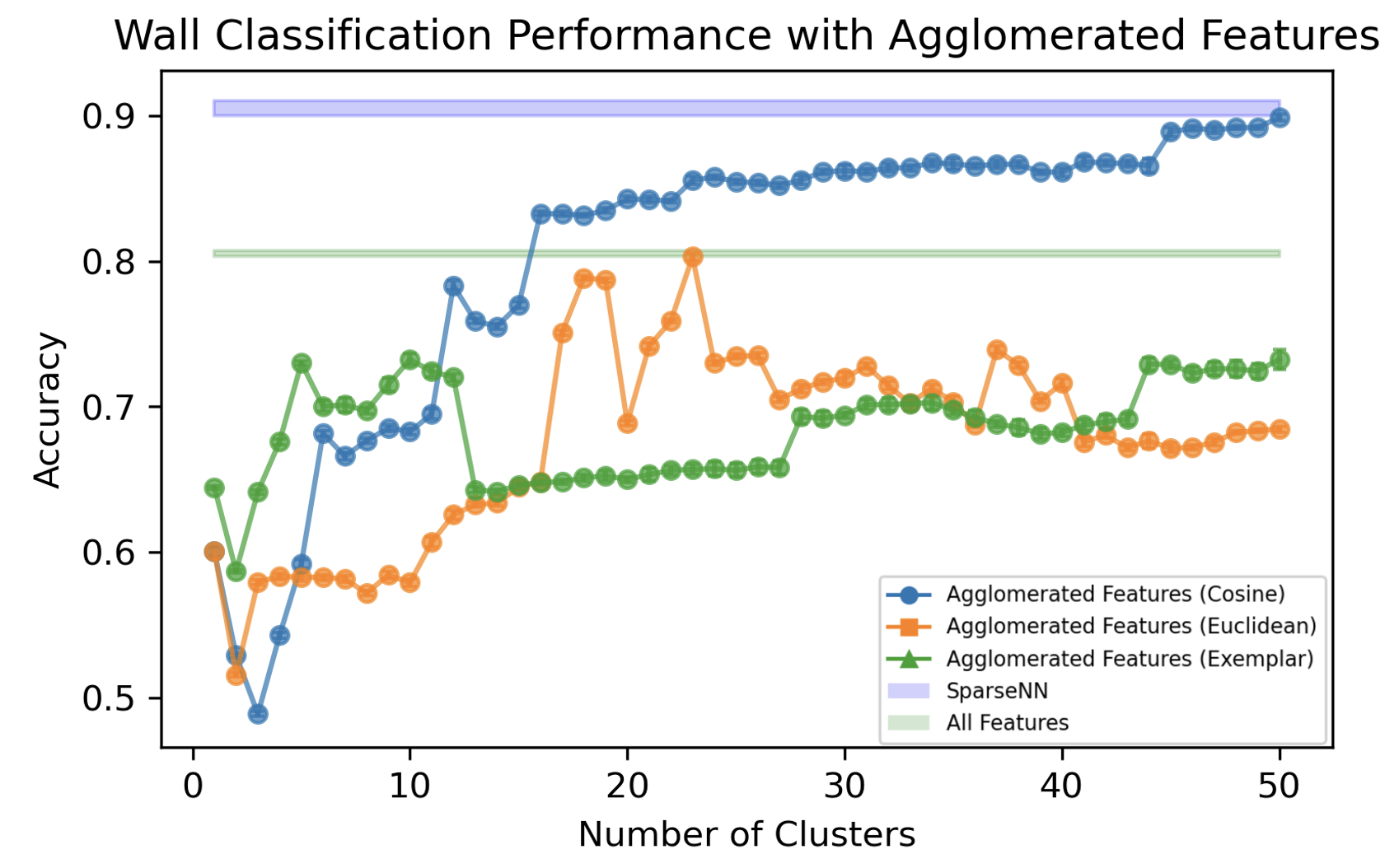}
    \caption{Agglomerated features selected using the cosine metric show substantially better performance for the RF model, with test accuracy approaching that of the SparseNN. The accuracy rises sharply up to around 18 clusters and.}
\label{fig:wall_classification_agglomerated_features_performance}
\end{figure}

As shown in Figure~\ref{fig:wall_classification_agglomerated_features_performance}, approximately 18 feature clusters are sufficient to match the prediction performance obtained using all 655 trace features, for both Euclidean and Cosine distance metrics. The SparseNN with an optimized architecture attains marginally higher accuracy overall.

\subsubsection{Performance Based Feature Selection: Permutation Feature Importance}

Analyzing the features independently by permutation feature importance (PFI) analysis reaffirms this hypothesis. Unlike the stud detection, features with significant positive importance for wall classification are mostly distributed between the 1 ns and 2.5 ns time signals. There are numerous features with smaller positive importance later in the trace, with an uptick in importance at around 6~ns, likely due to higher-order reflections that are important for the specific geometries of the envelopes in the study.

\begin{figure}[H]
    \centering
    \includegraphics{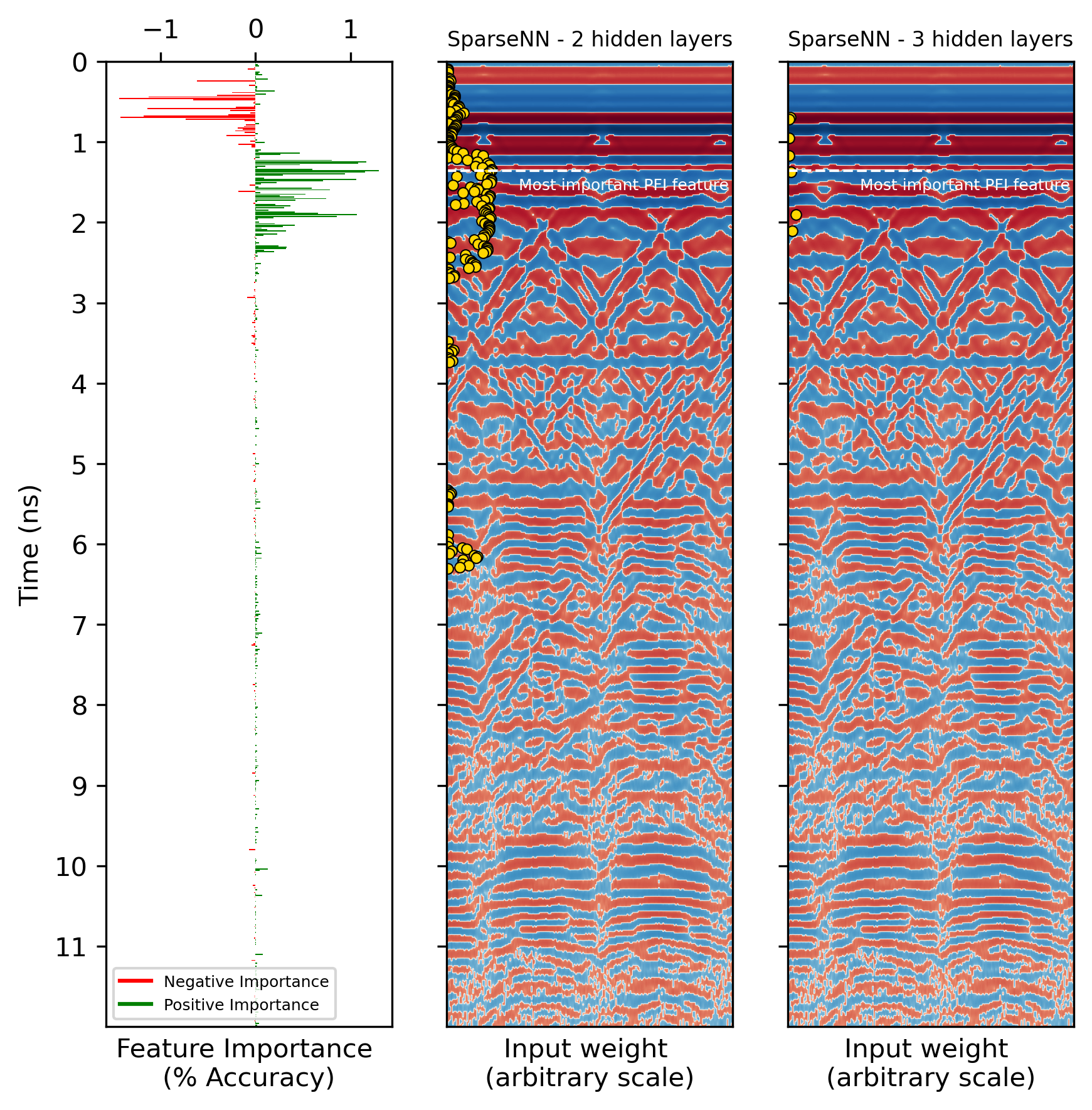}
    \caption{Feature importances for the wall classification task. Important features show strong correspondence with features selected by shallower, lightly trained SparseNNs. Both sets of features also include far-field signals that are likely higher-order signals not directly related to the wall type (~ 6 ns). However, deeper SparseNNs with more comprehensive training selected a much smaller number of features. They also disregarded far-field reflections, while retaining performance.}
    \label{fig:PFI_wall_classification}
\end{figure}

The PFI analysis revealed that only a small fraction of features meaningfully contribute to model performance. Notably, a subset of features located at early time indices (less than 1~ns) exhibited a strong negative impact, suggesting that their inclusion may degrade predictive accuracy. This motivates a more targeted approach to feature selection. While PFI provides a global ranking of individual feature importance, it does not account for interactions or redundancies among features. To address this limitation and identify a performance-optimized subset, Recursive Feature Elimination (RFE) is applied. RFE iteratively removes the least informative features while retraining the model at each step, allowing it to capture feature dependencies and converge toward a minimal, high-performing set.

\subsubsection{Recursive Feature Elimination}

A Recursive Feature Elimination with Cross-Validation (RFECV) is applied to the wall classification task, using a Random Forest classifier as the estimator and a stratified group k-fold strategy to preserve scan-level grouping during evaluation. The algorithm iteratively eliminates the least important features in steps of five, retraining the model at each stage, and selects the feature subset that yields the highest mean cross-validated accuracy. This is compared to the performance of the selected features on the unseen test set. 

\indent\subparagraph{Strategy 1: Stratified Cross Validation}\leavevmode\newline

The methodology for stratified cross-validation is identical to the process for stud detection, with the same train and test splits.

\begin{table}[H]
\centering
\begin{tabular}{ c c c }
\hline
\textbf{Configuration} & \textbf{Train} & \textbf{Validation} \\
\hline
1-5 & Segments of \texttt{I1} and \texttt{G3}  & Left-out segments of \texttt{I1} and \texttt{G3} \\
\hline
\end{tabular}
\caption{Train and validation set for different folds}  
\label{tab:wall_classification_stratified_cv_train_test_split}
\end{table}

\begin{figure}[H]
    \centering
    \includegraphics{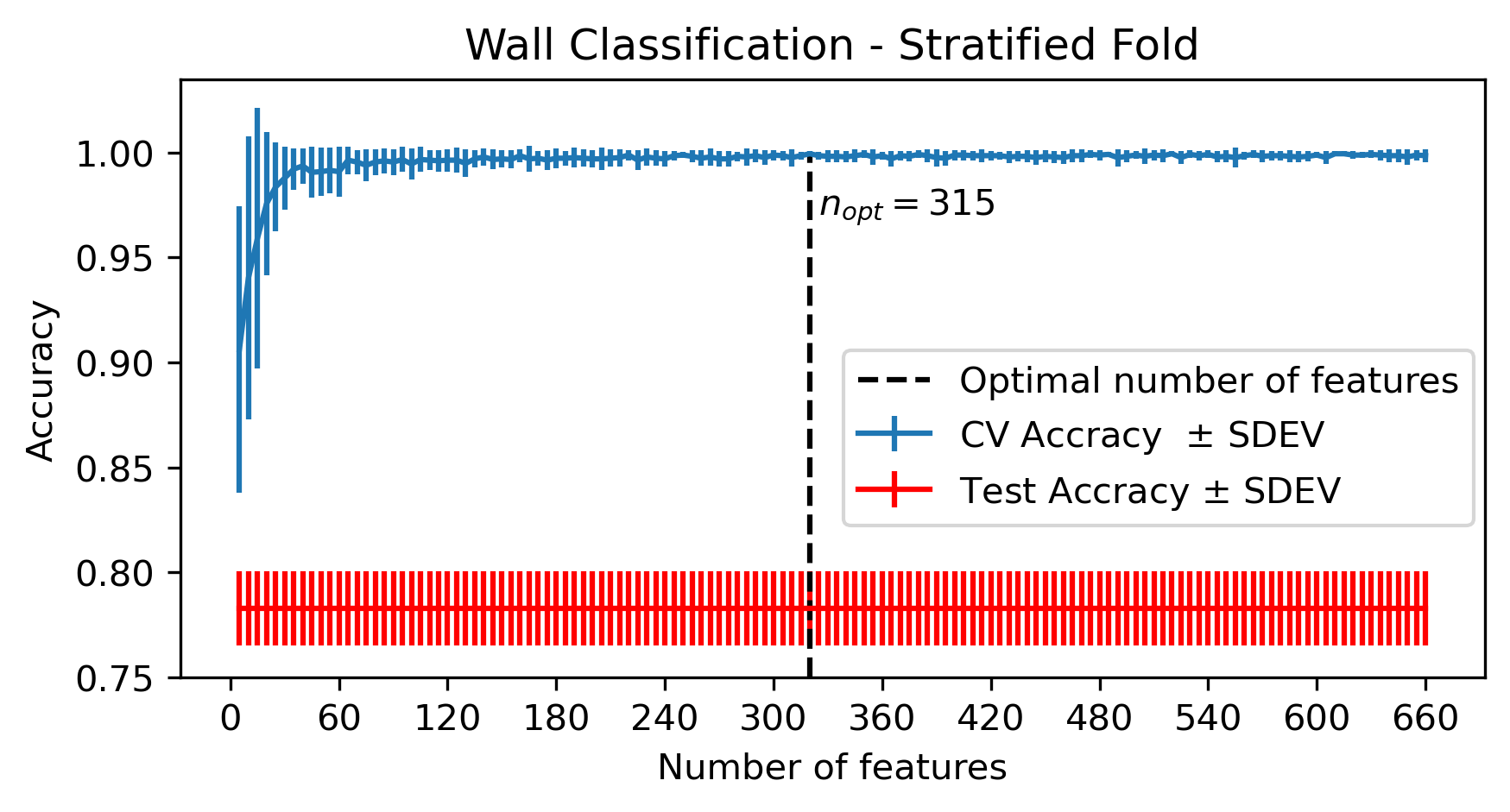}
    \caption{RFE with stratified CV for wall classification with minimal dataset (scans \texttt{I1} and \texttt{G3}). Performance close to optimum performance in the validation set (\(\sim\)1.0) is achieved using 48 features. The number of features required for optimum performance is 315. Accuracy for the test set (i.e. scans other than \texttt{I1} and \texttt{G3} is 0.78.}
    \label{fig:RFI_stratified_wall_classification}
\end{figure}

Test set predictions have a mean accuracy of 0.78, which is lower than the accuracy achieved using minimal sample (Scans \texttt{I1} and \texttt{G3}). This is to be expected because, as shown in Table~\ref{tab:wall_classification_stratified_cv_train_test_split}, the training process used segments of the same scans for the minimal sample case. Validation accuracy is \(\sim\)1.0.

\indent\subparagraph{Strategy 2: Stratified Group Cross Validation}\leavevmode\newline







\begin{table}[H]
\centering
\begin{tabular}{ c c c }
\hline
\textbf{Configuration} & \textbf{Train} & \textbf{Validation} \\
\hline
1 & 1,3,19 & 7 \\
\hline
2 & 1,7 & 3,19 \\
\hline
3 & 3,7,19 & 1 \\
\hline
\end{tabular}
\caption{Train and validation set for different folds}  
\label{tab:train_test_wall_background_classification}
\end{table}

\begin{figure}[H]
    \centering
    \includegraphics{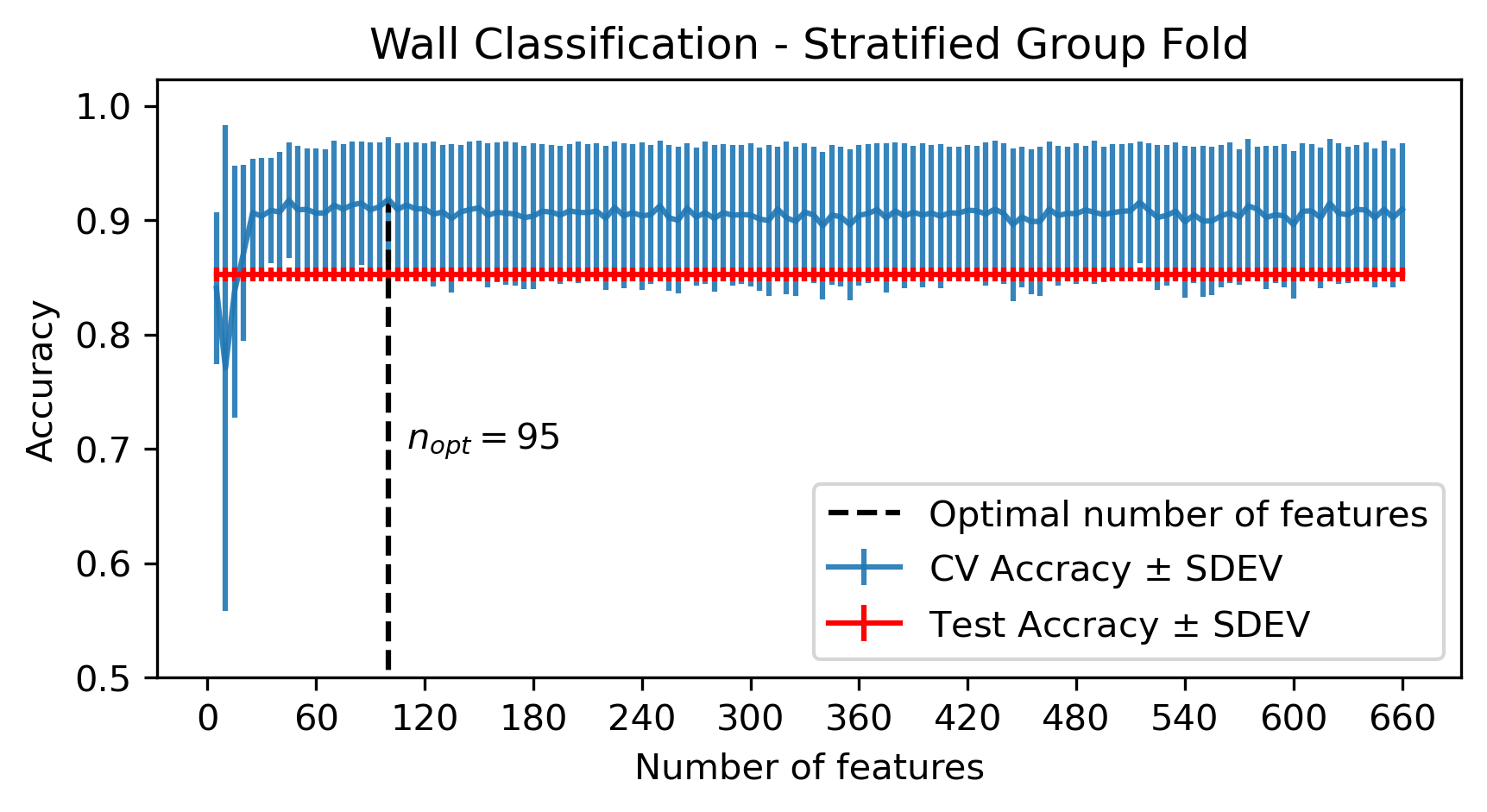}
    \caption{RFE with stratified grouped CV for wall classification. Close to optimum performance in the validation set (0.92 accuracy) is achieved using 16 features. The number of features required for optimum performance is 95. Accuracy for the test set (i.e. scans other than \texttt{I1} and \texttt{G3} is 0.85.). Using different}
    \label{fig:wall_rfe_grouped_performance}
\end{figure}

\indent\subparagraph{Sparse Neural Network}\leavevmode\newline

\begin{figure}[H]
    \centering  \includegraphics{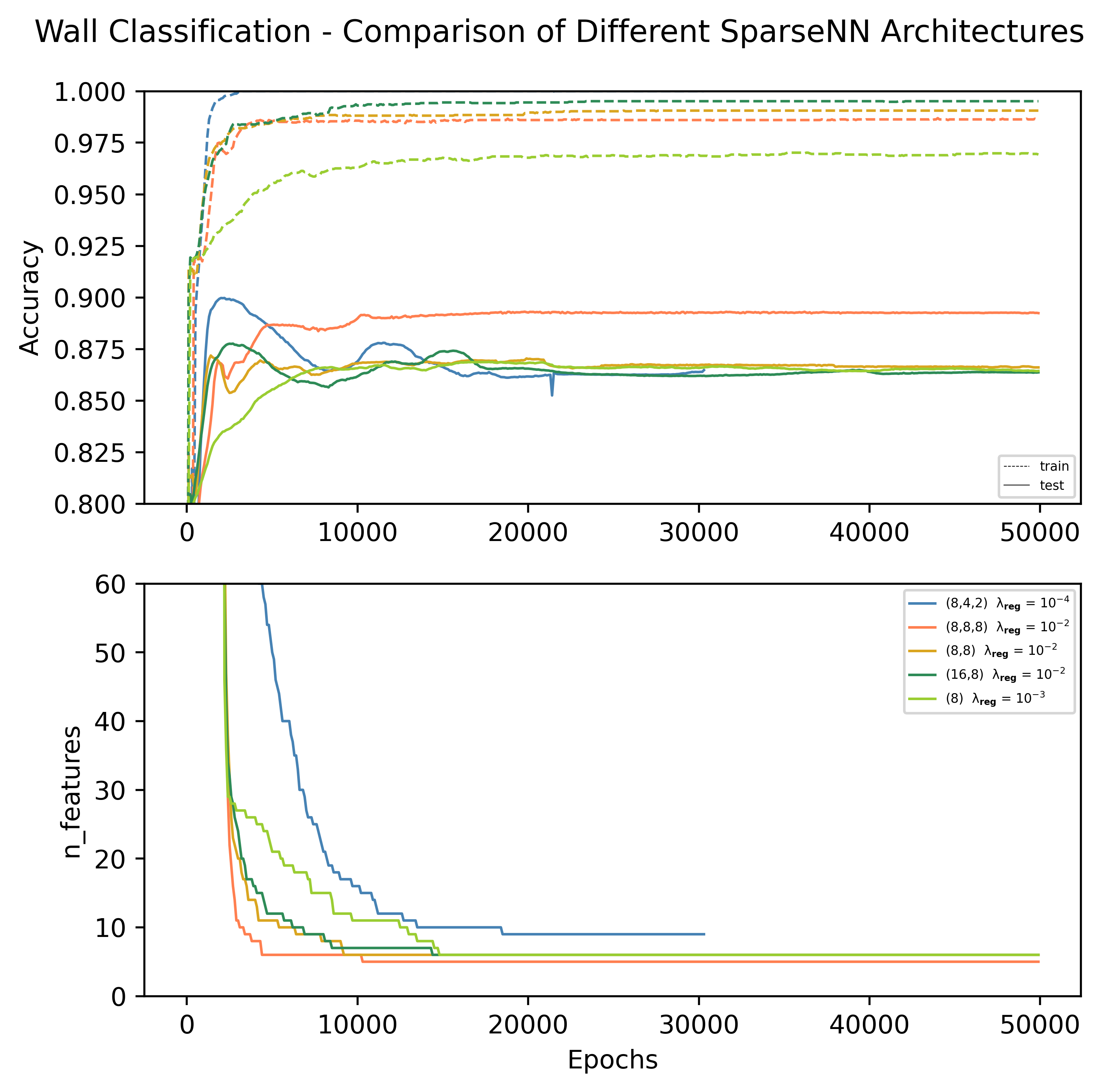}
    \caption{Wall classification performance for different SparseNN architectures and regularization loss coefficient combinations. The deepest SparseNN, with (8,8,8) hidden layers, produces the best test performance with the least number of input features. However, increasing the number of layers further did not produce appreciable performance improvement.}
\label{fig:wall_classification_SparseNN_performance}
\end{figure}

\section*{Acknowledgements}

This material is based upon work supported by the National Science Foundation under Grant IIS-2123343 and by the U.S. Department of Energy’s Office of Energy Efficiency and Renewable Energy (EERE) under the Building Technologies Office, Award Number DE-EE0009748.

\section*{Declaration of Generative AI and AI-Assisted Technologies}

During the preparation of this manuscript, the authors used ChatGPT (OpenAI) to assist with language editing and stylistic refinement. The authors reviewed, edited, and verified all content generated using this tool and take full responsibility for the accuracy, originality, and integrity of the published work.

\bibliographystyle{unsrt}
\bibliography{references.bib}

\end{document}